\documentclass[a4j,12pt]{report}
\usepackage[dvips]{graphicx}
\usepackage{color}
\usepackage{fancybox}
\begin{document}
\pagestyle{empty}

\title{\bf 
 \ \    \\
 \ \    \\
Formalism of Nonequilibrium Perturbation Theory \\  
 and Kondo Effect  
 \ \    \\
 \ \    \\
 \ \    \\
 \ \    \\
 \ \    \\
 \ \    \\
}
\author{\LARGE{\bf Mami Hamasaki}}

\maketitle

 \ \

 \ \

 \ \

{\huge \bf Abstract}

 \ \

 \ \

The formalism of nonequilibrium perturbation theory was constructed
 by Schwinger and Keldysh and then was developed with the diagrammatical
 technique by Lifshitz and Pitaevskii. Until now there has been
 widespread application to various researches in physics,  
condensed matter, plasmas, atoms and molecules, nuclear matter 
etc.. In spite of this, the formalism has not been 
established as perturbation theory. For example, 
there is no perturbative method to derive arbitrary 
self-energy properly. In addition, the connection with 
other formalism, e.g.,  the Matsubara imaginary-time 
perturbative formalism is uncertain. 
 Although there must be the relationship between 
self-energies in the perturbative formalism, 
such basic problems remain to be solved. 
 The solution is given by the present work. The
 real-time perturbative expansion is performed on the basis of the
 adiabatic theorem. As the results, the requirements of self-energies as
 functions in time are demonstrated and the formulated self-energies meet
 the known relations. Besides, it gives exact agreement with functions
 derived by perturbative expansion in imaginary-time and analytical
 continuity. As a consequence, it implies that the present formalism can be
 generalized. 

 \ \ 

Next, using the formulated self-energies, the behavior of
 the Kondo resonance is investigated for nonequilibrium states caused by
 bias voltage. As numerical results, the Kondo peak disappears when
 voltage exceeds the Kondo temperatures; it is supported by experiments
 for two terminal systems. 
 Over ten years, it has been being waited in expectation 
that the Kondo peak splits owing to bias voltage as a candidate 
for two channel Kondo effect. Nevertheless, 
it has not been observed in two terminal systems  
by experiments.  Here, it is discussed why the Kondo peak splitting  
may not arise in normal two terminal systems.

\tableofcontents


\chapter{Introduction}
\section{Nonequilibrium Perturbative Formalism \\
 ( Schwinger-Keldysh Formalism )} 

 \ \ 

The basic idea on the nonequilibrium perturbation theory 
 was proposed by Schwinger in 1961.[1] 
That included the essentials for the nonequilibrium perturbation 
theory. The idea of time reversal is the basis of the theory.   
The real time-contour has the positive and reverse time 
directions, so that it starts and ends at $t=-\infty$ by way of $t=\infty$. 
In addition, the matrix form in the nonequilibrium 
Green's functions is written after the real time-contour. 
After that, there are two main developments in the nonequilibrium 
perturbation theory.  One is the expansion of the formalism 
by means of the equation of motion in the Green's function
by Kadanoff and Baym[2]; there their time-contour includes    
imaginary-time path. Another is that the formalism was extended 
as the frame of the nonequilibrium perturbation theory 
by Keldysh in 1965.[3] The formalism is constructed 
using density matrix and S-matrix after the time-contour 
and the Dyson's equation in the matrix form 
of the nonequilibrium Green's functions 
on the basis of the idea of Schwinger.     
The Dyson's equation for the Keldysh formalism is given by 
\begin{eqnarray}
{\bf G}&=&  \ \  {\bf g} \ \  +  \ \ {\bf g}
 \ \ {\bf{\Sigma}} \ \ {\bf G},  
\end{eqnarray}
where 
\begin{eqnarray}
{\bf G}=\left[ \begin{array}
{ll}
G^{--} & G^< \\
 G^> & G^{++} \\
\end{array} 
\right],  \ \
{\bf {\Sigma}}=\left[ \begin{array}
{ll}{\Sigma}^{--} & {\Sigma}^< \\
{\Sigma}^> & {\Sigma}^{++} \\
\end{array} 
    \right].  \nonumber  
\end{eqnarray}

 \ \ 

Then the Keldysh formalism was studied 
in further detail and generalized    
as the formalism of the nonequilibrium Green's functions  
and the perturbative method 
with the help of the diagrammatic technique 
by Lifshitz and Pitaevskii.[4] 

\ \ 

The main progress in the formalism of the nonequilibrium Green's functions and 
the nonequilibrium perturbation theory is summarized with the aid of chart 
as follows:

 \ \ 

\shadowbox{ 
\begin{tabular}{c}
Basic idea on the nonequilibrium perturbation 
theory \\ ( the real time-contour and the matrix form  )  
was proposed \\ by Schwinger in 1961  \ \  \ \  \ \  \ \  \ \  \ \  \ \  \ \  \ \  \ \  \ \  \ \ 
 \ \  \ \  \ \  \ \  \ \  \ \  \ \  \ \  \ \  \ \  \ \  \ \ \end{tabular}
}

 \ \  \ \  \ \  \ \  \ \ \ \  \ \  \ \  \ \  \ \ {\huge ${ \downarrow}$}
  \ \  \ \  \ \  \ \  \ \   \ \  \ \  \ \    
  \ \  \ \  \ \  \ \  \ \   \ \  \ \  \ \    
\ \  \ \  \ \  \ \  \ \  \ \ {\huge ${ \downarrow}$}

\shadowbox{ 
\begin{tabular}{c}
Formalism was constructed \\ 
after the time-contour \\ 
and the matrix form 
\\ by Keldysh in 1965     
\end{tabular}
} \ \  \ \  \ \  \ \ 
\shadowbox{ 
\begin{tabular}{c}
Formalism was expanded \\ using the equation of motion \\ 
by Kadanoff and Baym \\ (1961-62)
 \ \  \ \  \ \  \ \ \ \ \ \ 
\end{tabular}
}

 \ \  \ \  \ \  \ \  \ \ \ \  \ \  \ \  \ \  \ \ {\huge ${ \downarrow}$}
  \ \  \ \  \ \  \ \  \ \   \ \  \ \  \ \    
  \ \  \ \  \ \  \ \  \ \   \ \  \ \  \ \    
\ \  \ \  \ \  \ \  \ \  \ \ {\huge ${ \downarrow}$}

 \ \  \ \  \ \  \ \  \ \ \ \  \ \  \ \  \ \  \ \ 
  \ \  \ \  \ \  \ \  \ \   \ \  \ \  \ \    
  \ \  \ \  \ \  \ \  \ \   \ \  \ \  \ \    
\ \  \ \  \ \ \ovalbox{ to be continued!} 

\shadowbox{ 
\begin{tabular}{c}
Formalism was developed \\  
with the diagrammatical \\ technique 
by Lifshitz \\ and Pitaevskii   
\end{tabular}
}

 \ \  \ \  \ \  \ \  \ \ \ \  \ \  \ \  \ \  \ \ {\huge ${ \downarrow}$}
  \ \  \ \  \ \  \ \  \ \   \ \  \ \  \ \    
  \ \  \ \  \ \  \ \  \ \   \ \  \ \  \ \    
\ \  \ \  \ \  \ \  \ \  \ \

 \ \  \ \  \ \  \ \  \ \ \ \  \ \  \ \  \ \  \ \ {\huge ${ \downarrow}$}
  \ \  \ \  \ \  \ \  \ \   \ \  \ \  \ \    
  \ \  \ \  \ \  \ \  \ \   \ \  \ \  \ \    
\ \  \ \  \ \  \ \  \ \  \ \

 \ \  \ \  \ \  \ \  \ \ \ \  \ \  \ \  \ \  \ \ {\huge ${ \downarrow}$}
  \ \  \ \  \ \  \ \  \ \   \ \  \ \  \ \    
  \ \  \ \  \ \  \ \  \ \   \ \  \ \  \ \    
\ \  \ \  \ \  \ \  \ \  \ \

 \ \  \ \   \shadowbox{\large \bf the present work } 
  \ \  \ \  \ \  \ \  \ \   \ \  \ \  \ \    
  \ \  \ \  \ \  \ \  \ \   \ \  \ \  \ \    
\ \  \ \  \ \  \ \  \ \  \ \

 \ \  \ \  \ \  \ \  \ \ \ \  \ \  \ \  \ \  \ \ {\huge ${ \downarrow}$}
  \ \  \ \  \ \  \ \  \ \   \ \  \ \  \ \    
  \ \  \ \  \ \  \ \  \ \   \ \  \ \  \ \    
\ \  \ \  \ \  \ \  \ \  \ \

 \ \  \ \  \ \  \ \  \ \ \ovalbox{ to be continued!} 
  \ \  \ \  \ \  \ \  \ \   \ \  \ \  \ \    
  \ \  \ \  \ \  \ \  \ \   \ \  \ \  \ \    
\ \  \ \  \ \  \ \  \ \  \ \ 

 \ \ 

The connections between these ways have also been investigated   
 and those have been confirmed in accordance together.  

 \ \ 

Recently, the formalism of the nonequilibrium Green's functions 
and the nonequilibrium perturbative method have been applied widely 
to the various fields: condensed matter, plasmas, atoms 
and molecules, nuclear matter etc..[5]  
Especially,  the application to the problems of mesoscopic systems: 
the transports in quantum dots and quantum wire, 
has worked with great success[6-8]; for instance,      
electrical current  and current noise are expressed 
in terms of the nonequilibrium Green's functions. 
For current, 
\begin{eqnarray}
2{\langle}{\hat{J}}_{\sigma}(i-1,i){\rangle}\, ={\frac{2eW}{h}}\int{dE} 
 \bigl[\, G^<_{i-1,i{\sigma}}(E)-G^<_{i,i-1{\sigma}}(E)\, \bigr], 
\end{eqnarray}
where $W$ denotes the hopping matrix element. 
This reduces to the Landauer formula:[9] 
\begin{eqnarray}
2{\langle}\hat{J}{\rangle}=
{\frac {2e}{h}}\int dE [f_L(E)-f_R(E)]{\cal T}(E), 
\end{eqnarray}
where $f_L$ and $f_R$ are the Fermi distribution functions 
in the isolated left and right leads, respectively. 
For the current noise at zero-frequency 
from the autocorrelation function of the current[10]:
\begin{eqnarray}
2S^0_{{\sigma}{\sigma}}(i-1,i)=
-2\left({\frac{2e^2W^2}{h}}\right)\int dE 
 [ G^{<}_{i,i-1{\sigma}}(E)\, 
G^{>}_{i,i-1{\sigma}}(E) & &   \nonumber \\
 -G^{<}_{i-1,i-1{\sigma}}(E)\, G^{>}_{i,i{\sigma}}(E)& &  \nonumber \\
 -G^{<}_{i,i{\sigma}}(E)\, G^{>}_{i-1,i-1{\sigma}}(E)& &  \nonumber \\
 +G^{<}_{i-1,i{\sigma}}(E)\, G^{>}_{i-1,i{\sigma}}(E)& &  ]. 
\end{eqnarray}
 It reduces to the Khlus-Lesovik formula for one channel at zero-frequency 
for shot noise:[11]  
\begin{eqnarray}
S^{KL}=2\left({\frac {2e^2}{h}}\right)\int dE 
&[&f_R(E)\{1-f_R(E){\}}{\cal T}(E)\nonumber \\
&+&f_L(E)\{1-f_L(E)\}{\cal T}(E)\nonumber \\
&+&\{f_L(E)-f_R(E)\}^2 
{\cal T}(E)\{1-{\cal T}(E)\}]. 
\end{eqnarray}
Here, the transmission probability  
through a noninteracting system in Eqs. ( 1.3 ) and ( 1.5 ) 
is written by  
\begin{eqnarray}
{\cal T}(E)&=&{\Gamma}_L(E)G^r_{1n{\sigma}}(E)
{\Gamma}_R(E){G}^a_{n1{\sigma}}(E).  
\end{eqnarray}
${\Gamma}_{L}$ and ${\Gamma}_{R}$ 
mean the coupling functions with the left and the right leads, 
respectively, and  $G^r$ and ${G}^a$ are 
retarded and advanced Green's functions, severally,   
as described in detail later.
The expression for current noise also reduces 
to the Johnson-Nyquist noise for thermal noise: 
$4{\cal G}k_BT$  ( 
here, ${\cal G}$ signifies conductance, and 
 $k_B$ is the Boltzmann's constant 
and $T$ denotes temperature ).[12] 
These are excellently compatible with the values observed 
by experiments. 

 \ \ 

Nonetheless, the formalism of the nonequilibrium 
perturbation theory has not been completed yet; 
the perturbative methods with the diagrammatic technique  
remain to be clarified well. 
In particular, despite of that the definition of the nonequilibrium Green's functions   
 is given in time, the real-time perturbative expansion 
on the adiabatic theorem[13]$-$the basics of method of 
the perturbative expansion$-$is almost unknown.  
The general formalism to formulate arbitrary self-energy 
has still not been established and the connection 
with other formalism, for instance,  
the Matsubara imaginary-time perturbative formalism[14]  
is obscure. For this reason, the present work succeeds 
to the Keldysh formalism and the diagrammatical method 
of Lifshitz and Pitaevskii to make progress in 
the formalism of the real-time perturbative expansion 
based on the adiabatic theorem.  
If this matrix form equation Eq. ( 1.1 ) works exactly 
as the Dyson's equation,   
this equation must be convertible into 
the Dyson's equations for retarded and advanced    
Green's functions. Accordingly,  
there must be the relations between self-energies. 
However, since the definition of 
the self-energies for this Dyson's equation  
in matrix form is not given, 
the self-energies drawn from perturbative expansion 
have not been made clear. 

In the present work, thus, the solution on the relations 
between self-energies is given. 
Using the solution, the retarded and advanced self-energies 
 are derived from the self-energies in matrix form. 
Then, the derived retarded and advanced self-energies 
 meet  the conditions required as functions in time 
and the generally known relations on nonequilibrium 
 Green's functions are fulfilled. 
Additionally, the formulated self-energies are 
in agreement with those derived by perturbative expansion 
in imaginary-time and analytical continuity.       
Thereby, the solution is in accordance with the generally known 
formation. As a result, it infers that the formalism 
of the nonequilibrium perturbation theory can be generalized.  

 \ \ 

\section{Kondo Effect} 

 \ \ 

The Kondo effect was discovered forty years ago[15]; 
the phenomenon of the minimum of the electrical 
resistivity in metals was explained 
in view of the interaction between conduction electrons 
and impurity by Kondo. After that, the Kondo physics 
was clarified from Landau's Fermi liquid theory[16], 
the renormalization group[17] and scaling[18]. 
Besides, the generalized Kondo problem  
 with more than one channel or one impurity was  
proposed.[19] It has then been investigated in further detail.[20-23]
Especially, the resistivity has 
been expressed for the multichannel Kondo effect 
 by the conformal field theoretical work,[20-22] 
in agreement with experiments, 
as mentioned in Chapter 4. 

Moreover, the Kondo effect in electron transport through
 a quantum dot was predicted theoretically
 at the end of 1980s[24-26]. After a decade, finally, 
this phenomenon was observed.[27] 
The Kondo effect was studied theoretically by use of the 
Anderson model. From scaling theory[18],  
the Kondo temperatures 
\begin{eqnarray}
k_BT_K{\sim} D
e^{-{\pi}U/8{\Gamma}}.
\end{eqnarray}
Here $D$ means the band-width and  
the Coulomb interaction $U$, 
 and ${\Gamma}$ is  
the coupling function with leads,  
 corresponding to the density of states 
for conduction electron. 
The Kondo temperatures correspondent to the strength of 
the Kondo coupling decrease 
with increasing $U$, as found from Eq. ( 1.7 ). 
 The predictions from theoretical work 
using Anderson model were confirmed experimentally. 
In the Kondo regime, the conductance was observed to reach 
the unitarity limit and the Kondo temperatures estimated 
 from observation[28] are in excellent  agreement 
with the expression  derived by the use of 
scaling theory for the asymmetric 
Anderson model[29]:
\begin{eqnarray}
T_K={ \frac { \sqrt{{\Gamma}U} }{2} }
e^{{\pi}E_0(E_0+U)/{\Gamma}U}
\end{eqnarray} 
where on-site energy is $E_0$.   
The perturbative approach, 
the Yamada-Yosida theory[30]$-$the perturbation 
theory for equilibrium based on  the Fermi liquid theory[16] 
with the  Matsubara imaginary-time perturbative method[14] 
 is quite successful. 

Furthermore, the Kondo effect in a quantum dot was studied 
 for nonequilibrium system where bias voltage is applied.[31] 
We have to know not only the Kondo effect but also 
the nonequilibrium state caused by bias voltage. 
The Yamada-Yosida theory was extended to nonequilibrium 
systems with the help of the Keldysh formalism     
 and the Kondo effect in nonequilibrium system  
was studied. As the results, it was shown that 
for bias voltage higher than the Kondo temperatures, 
the Kondo resonance disappears in the spectral function 
with the second-order self-energy of the Anderson model.[8]
This results have been discussed little.     
After that, on experiments, it has been observed 
that the Kondo effect is suppressed when 
source-drain bias voltage is comparable to or exceeds 
the Kondo temperatures.[32,33] 
The numerical results of the present work are also consistent with 
those.  For the Kondo effect in nonequilibrium systems, 
it has been expected that the Kondo peak splits by bias voltage     
and that the two separated energy levels made in a quantum dot  
by the Kondo coupling act as two channels 
for two channel Kondo effect.  
In order to search that, every efforts have been done for many years.  
 Such the phenomenon has however, never been observed for a quantum dot connected 
with two normal leads. In the present paper, 
it is discussed the reason why the Kondo peak is just broken 
and the Kondo peak splitting may not take place 
in simple two terminal systems with leads of the continuous energy states. 

\chapter{ Nonequilibrium  Perturbation Theory}

 \ \ 

A thermal average can be gained  
on the basis of the nonequilibrium perturbation 
 theory.[1,3,4,13,34-37] 

The perturbation theory is based on 
the adiabatic theorem ( called the Gell-Mann and Low's 
theorem[13] ). 
The Hamiltonian is given by 
\begin{eqnarray}
{\cal H}={\cal H_{\rm 0}}+{\cal H_{\rm I}}. 
\end{eqnarray}
${\cal H_{\rm 0}}$ and ${\cal H_{\rm I}}$ 
 are unperturbed and perturbed terms, respectively.  
Here, assuming that we can know only the state of ${\cal H_{\rm 0}}$ 
at $t=-{\infty}$, thereby, initially ${\cal H_{\rm I}}=0$ at $t=-{\infty}$,    
 so that the system is equilibrium and/or noninteracting state. 
The perturbation is turned on at $t=-{\infty}$     
and introduced adiabatically. Then the perturbation is brought wholly 
into the system at $t=0$; around $t=0$, the system is 
regarded as stationary nonequilibrium and/or interacting state. 
After that, the perturbation  
is taken away adiabatically and disappears at $t={\infty}$. 
When the time evolution of the state is reversible,
the state at $t={\infty}$ can be expressed 
using the state at $t=-{\infty}$ by adding the phase factor.       

 \ \ 

Now, let us consider the nonequilibrium state. 
If the time evolution of the state is 
irreversible for the nonequilibrium state, then, 
the state at $t={\infty}$ is  not well-defined;  
when the perturbation is removed entirely at $t={\infty}$,
the state does not come back to the same state 
 as at $t=-{\infty}$. 
In this case, the ordinary perturbative method 
should be improved; the time evolution should return to  
the well-defined state at $t=-{\infty}$. 
Accordingly, the time evolution is performed 
along the real-time contour which starts and ends 
at $t=-{\infty}$ as illustrated in Fig. 2.1. 
It is the extension of 
the Gell-Mann and Low's theorem.[13]   

\begin{figure}[htbp]
\begin{center}
\includegraphics[width=7.0cm]{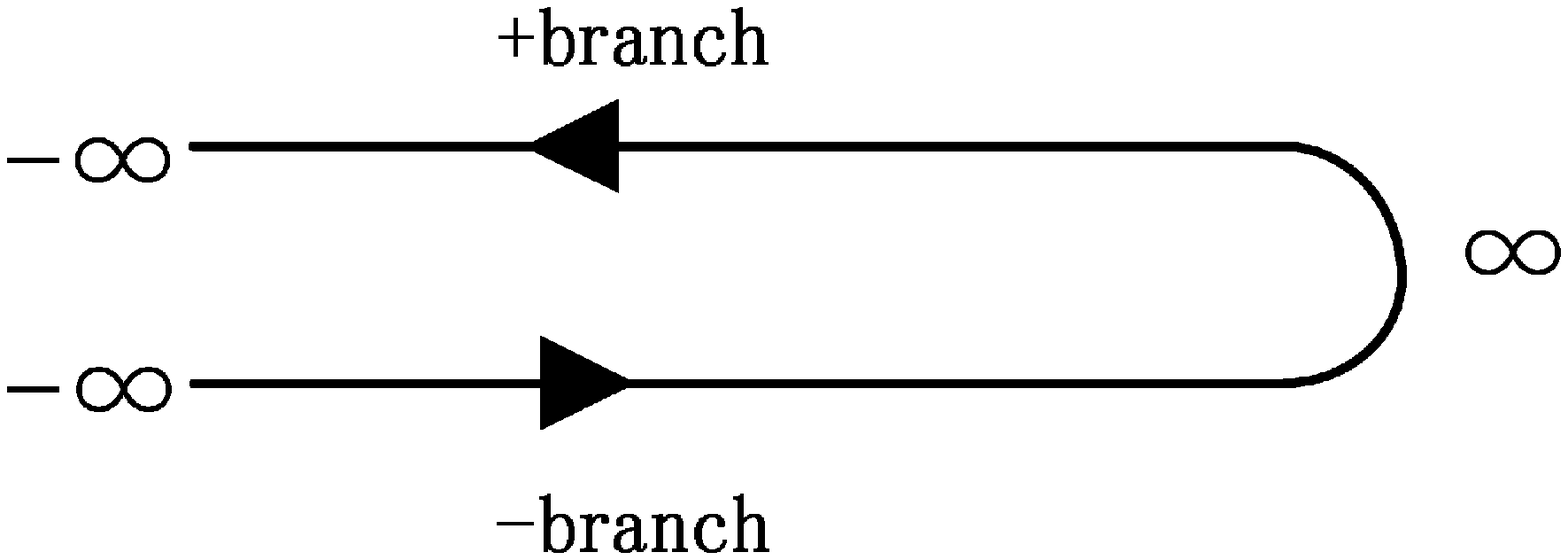}
\caption{time contour which starts and ends at  
$t=-\infty$ through $t=\infty$}
\label{Fig1}
\end{center}
\end{figure}

 \ \ 

 \ \ 

 \ \ 

\section{S-matrix (S-operator)}

 \ \ 

The Hamiltonian is given by Eq. ( 2.1 ). 
The time evolution in the interaction representation 
is expressed in terms of S-matrix by 
\begin{eqnarray}
{\tilde{\psi}}(t)={S(t,t_{\rm 0})}{\tilde{\psi}}(t_{\rm 0}).
\end{eqnarray}
S-matrix ${S(t,t_{\rm 0})}$ is defined by   
\begin{eqnarray}
{S(t,t_{\rm 0})}=e^{i{\cal {H}_{\rm 0}}t/ {\hbar}}
e^{-i{\cal {H}}(t-t_0)/ {\hbar}} 
e^{-i{\cal {H}_{\rm 0}}t_{\rm 0}/{\hbar}}, 
\end{eqnarray}
and has the following properties:  
\begin{eqnarray}
S(t_{\rm 0},t_{\rm 0})&=&1,  \\
S(t,t_{\rm 0})^{\dag}&=&S(t_{\rm 0},t),   \\
S(t_1,t_2)S(t_2,t_3)&=&S(t_1,t_3),  \\
i{\hbar}{\frac{ {\partial}S(t,t_{\rm 0})}{ {\partial} t }}
&=&{\tilde{\cal {H}}_{\rm I}}S(t,t_{\rm 0}).
\end{eqnarray}
Equation ( 2.7 ) can be solved formally by      
\begin{eqnarray}
{\cal S}(t,t_{\rm 0})&=&1+\sum_{n=1}^{\infty}
{ \frac {1}{n!} }\left({ \frac{-i}{\hbar} }\right)^n
\int_{t_{\rm 0}}^t dt_1 {\ldots} 
\int_{t_{\rm 0}}^t dt_n 
{\rm T}\left[{\tilde{\cal {H}}_{\rm I}}(t_1){\ldots}
{\tilde{\cal {H}}_{\rm I}}(t_n)\right] \nonumber   \\
&=&{\rm T}\left[ {\exp} \left\{ { \frac{-i}{\hbar} }\int_{t_{\rm 0}}^t 
dt^{'}{\tilde{\cal {H}}_{\rm I}}(t^{'}) \right\} \right],   \\ 
{\cal S}(t,t_{\rm 0})^{\dag}&=&{\cal S}(t_{\rm 0},t)  
={\tilde{\rm T}}\left[ {\exp} \left\{ { \frac{i}{\hbar} }
\int_{t_{\rm 0}}^t dt^{'}{\tilde{\cal {H}}_{\rm I}}(t^{'}) \right\} \right].  
\end{eqnarray}
Here, the time ordering operator ${\rm T}$ arranges in 
chronological order and ${\rm \tilde{T}}$ is the anti time 
ordering operator which arranges in the reverse of 
chronological order. 
For the time evolution along the time contour as in Fig. 2.1,
S-matrices, Eqs. (2.8) and (2.9)
are required for paths on the $(-)$ branch from $t=-{\infty}$ to $t={\infty}$  
and on the $(+)$ branch from $t={\infty}$ to $t=-{\infty}$, respectively.  

 \ \ 

In the same way, S-matrix for imaginary-time perturbative 
formalism is defined by 
\begin{eqnarray}
{S_{\tau}({\tau},{\tau}_{\rm 0})}=e^{{\cal {H}_{\rm 0}}{\tau}/ {\hbar}}
e^{-{\cal {H}}({\tau}-{\tau}_0)/ {\hbar}} 
e^{-{\cal {H}_{\rm 0}}{\tau}_{\rm 0}/{\hbar}} 
\end{eqnarray}
and is also written by 
\begin{eqnarray}
{\cal S}_{\tau}({\tau},{\tau}_{\rm 0})&=&1+\sum_{n=1}^{\infty}
{ \frac {1}{n!} }\left({ \frac{-1}{\hbar} }\right)^n
\int_{{\tau}_{\rm 0}}^{\tau} d{\tau}_1 {\ldots} 
\int_{{\tau}_{\rm 0}}^{\tau} d{\tau}_n 
{\rm T_{\tau}}\left[{\tilde{\cal {H}}_{\rm I}}({\tau}_1){\ldots}
{\tilde{\cal {H}}_{\rm I}}({\tau}_n)\right] \nonumber   \\
&=&{\rm T}_{\tau}\left[ {\exp} \left\{ { \frac{-1}{\hbar} }\int_{{\tau}_{\rm 0}}^{\tau} 
d{\tau}^{'}{\tilde{\cal {H}}_{\rm I}}({\tau}^{'}) \right\} \right].    
\end{eqnarray}
These are the requisites for the Matsubara imaginary-time 
perturbative formalism.[14] 

 \ \ 

\section{Matsubara Imaginary-Time 
Perturbative Formalism }

 \ \ 

For thermal equilibrium, the statistical operator ( density matrix ) is 
written in Gibbs form for the grand canonical ensemble by 
\begin{eqnarray}
{\varrho}_{G}={\frac{e^{-{\beta}({\cal H}-{\mu}N)}}
{{\rm Tr}e^{-{\beta}({\cal H}-{\mu}N)}}}=e^{{\beta}({\Omega}-{\cal H}+{\mu}N)}.
\end{eqnarray}
${\beta}=1/k_BT.$ 
By rearranging Eq. ( 2.10 ), we have 
\begin{eqnarray}
e^{-{\beta}({\cal H}-{\mu}N)}= 
e^{-{\beta}({\cal H_{\rm 0}}-{\mu}N)}{S_{\tau}({\beta}{\hbar},0)}. 
\end{eqnarray}
By substitution of Eq. ( 2.13 ) into Eq. ( 2.12 ), 
the thermal average for equilibrium is obtained by  
\begin{eqnarray}
{\langle}{\ldots}{\rangle}=
{\frac{{\rm Tr}[e^{-{\beta}({\cal H}-{\mu}N)}{\ldots}]}
{{\rm Tr}[e^{-{\beta}({\cal H}-{\mu}N)}]}}
={\frac{{\rm Tr}[e^{-{\beta}({\cal H_{\rm 0}}-{\mu}N)}{S_{\tau}({\beta}{\hbar},0)}{\ldots}]}
{{\rm Tr}[e^{-{\beta}({\cal H_{\rm 0}}-{\mu}N)}{S_{\tau}({\beta}{\hbar},0)}].}} 
\end{eqnarray}
By insertion of Eq. ( 2.11 ) in Eq. ( 2.14 ),  
the perturbative expansion is executed 
for the Matsubara imaginary-time perturbative formalism[14] 
using the Bloch and De Dominicis's theorem[38-41]. 
The Matsubara Green's function is defined by  
\begin{eqnarray}
G({\tau}){\equiv}-{\langle}{\rm T}_{{\tau}}\hat{d}({\tau})
 \hat{d}^{\dag}(0){\rangle}.
\end{eqnarray}
The functions in terms of the Matsubara Green's function 
in imaginary-time are converted 
into those in the Matsubara frequency 
by the Fourier transformation for the Matsubara Green's function:
\begin{eqnarray}
G(i{\omega}_n)=\int_{0}^{{\beta}} d{\tau}e^{i{\omega}_n{\tau}}G({\tau}). 
\end{eqnarray}
The Matsubara frequency,  
${\omega}_n{\equiv}$
$(2n+1){\pi}/{\beta}$ 
for fermion and ${\omega}_n{\equiv}$
$2n{\pi}/{\beta}$ 
for boson,  $( n=0,{\pm}1,$ ${\pm}2,$ ${\pm}3$ ${\ldots}  \ \ )$;  
thereby, the Matsubara Green's function is periodic. 

 \ \ 

After that, the analytical continuity is performed. 
For fermion, when the Taylor expansion for the function 
${e^{{\beta}z}+1}$ around poles $z=z^0$ on imaginary axis is done, 
then, ${e^{{\beta}z^0}+1}=0$, hence, 
$z^0=(2n+1){\pi}i/{\beta}$, and  
as approximation, ${e^{{\beta}z}+1}{\approx}{\beta}e^{{\beta}z^0}(z-z^0)$. 
The residue theorem yields 
the conversion of sum of the functions in the Matsubara frequency into 
the contour integral by 
\begin{eqnarray}
{\frac{1}{\beta}}{\sum_l}g(i{\omega}_l)=
{\frac{1}{2{\pi}i}}\int_{C}{\frac{g(z)}{e^{{\beta}z}+1.}}
\end{eqnarray}
It should be noted that the contour $C$ in integral surrounds 
the poles $z^0$ on imaginary axis.  
After that, for example, 
when $g(z)=1/(z-{\epsilon}_d$), then, 
the contour integral with contour $C'$ enclosing 
${\epsilon}_d$, a pole on real axis is executed by   
\begin{eqnarray}
{\frac{1}{2{\pi}i}}\int_{C'}{\frac{g(z)}{e^{{\beta}z}+1}}
=f({\epsilon}_d), 
\end{eqnarray}
where $f({\epsilon}_d)$ is the Fermi distribution function. 

In the same way, in boson case,  for poles $z=z^0$ on imaginary axis, 
${e^{{\beta}z^0}-1}=0$, i.e. $z^0=2n{\pi}i/{\beta}$,   
 and approximately, ${e^{{\beta}z}-1} {\approx}  {\beta}e^{{\beta}z^0}(z-z^0)$, 
 we have 
\begin{eqnarray}
{\frac{1}{\beta}}{\sum_l}g(i{\omega}_l)=-
{\frac{1}{2{\pi}i}}\int_{C}{\frac{g(z)}{e^{{\beta}z}-1.}}
\end{eqnarray}

 \ \ 

Then, by changing the contour integral around poles 
on imaginary axis into the contour integral parallel with 
real axis ( e.g. $E{\pm}i{\delta}$ ), 
we have the functions written in terms of retarded and advanced 
 Green's functions in energy. 
For high-order perturbation theory, 
the analytical continuation is so complicated. 
As the method of analytical continuation, {\'{E}}liashberg's 
method[42] is known.[4,41]  

 \ \ 

\section{Nonequilibrium Perturbative Formalism 
 }
\subsection{Nonequilibrium Real-Time Perturbative Formalism 
 }

 \ \ 

For nonequilibrium, Equation ( 2.12 ) is not exact.  
We should note that there are no specific limitations 
upon the statistical operator. 
The von Neumann's statistical operator is expressed 
independently of whether the states are at thermal equilibrium
or nonequilibrium by[35,37]
\begin{eqnarray}
{\varrho}_S(t)=\sum_{m}|m_S(t){\rangle}P_m{\langle}m_S(t)| 
\end{eqnarray}
in the Schr{\"o}dinger representation.  
Here, $P_m$ is probability that the system is in  
 state $m$ and $|m_S(t){\rangle}$ is the state in the Schr{\"o}dinger 
representation.    
It satisfies the Liouville 
equation by 
\begin{eqnarray}
i{\hbar}{\frac{ {\partial}{\varrho}_S }{ {\partial} t }}
=[{\cal {H}},{\varrho}_S]. 
\end{eqnarray}
The statistical operator in the interaction representation is given by  
\begin{eqnarray}
{\tilde{\varrho}(t)}&=&e^{i{\cal {H}_{\rm 0}}t/ {\hbar}}
{\varrho}_S(t) e^{-i{\cal {H}_{\rm 0}}t/{\hbar}},      
\end{eqnarray}
and also obeys the Liouville equation by  
\begin{eqnarray}
i{\hbar}{\frac{ {\partial}{\tilde{\varrho}} }{ {\partial} t }}
&=&[{\tilde{\cal {H}}}_{\rm I},{\tilde{\varrho}}]. 
\end{eqnarray}
As a matter of course, 
\begin{eqnarray}
{\varrho}_S(0)={\varrho}_H(0)={\tilde{\varrho}(0)},   
\end{eqnarray}
 where ${{\varrho}_H}(t)$ is statistical operator 
in the Heisenberg representation. 
The time evolution is written by means of S-matrix by 
\begin{eqnarray}
{\tilde{\varrho}}(t)=S(t,t_{\rm 0}){\tilde{\varrho}}(t_0)S(t_{\rm 0},t).
\end{eqnarray}
These are the properties of the von Neumann's statistical operator. 
Although the von Neumann's statistical operator Eq. ( 2.20 ) is 
not an explicit expression, it is still considered 
to be the same type as Eq. ( 2.14 ) so as  
 to execute the perturbative expansion.      

 \ \

 \ \ 

The thermal average for nonequilibrium is drawn   
in view of the analogy with 
the imaginary-time perturbative method  
for the Matsubara Green's function 
using S-matrix and the time evolution of the statistical operator, 
Eq.( 2.25 ). Thus the thermal average of the operators in 
the Heisenberg representation 
 at $t=0$ can be brought, for example by[3,4,34,37] 
\begin{eqnarray}
 & & {\langle}{\rm T} A(t)B(t^{'}){\rangle}  \nonumber   \\
&{\equiv}&{\rm Tr} \left[ {\varrho}_H(0){\rm T}A(t)B(t^{'})\right]  
\nonumber   \\ 
&=&{\rm Tr} \left[ {\tilde{\varrho}}(-\infty)
{\cal S}(-\infty,0){\rm T}A(t)B(t^{'}){\cal S}(0,-\infty)\right]
\nonumber   \\
&=&{\rm Tr} \left[ {\tilde{\varrho}}(-\infty)
{\cal S}(-\infty,\infty)\left\{{\rm T}{\cal S}(\infty,-\infty)
{\tilde{A}(t^-)}{\tilde{B}(t^{'-})}\right\}\right]
\nonumber  \\ 
&=&\sum_{n=1}^{\infty}\sum_{m=1}^{\infty}
{ \frac {1}{n!} }{ \frac {1}{m!} }
\left({ \frac{i}{\hbar} }\right)^n
\left({ \frac{-i}{\hbar} }\right)^m
\int_{-\infty}^{\infty} dt_1 {\ldots} 
\int_{-\infty}^{\infty} dt_n 
\int_{-\infty}^{\infty} dt_1^{'} {\ldots} 
\int_{-\infty}^{\infty} dt_m^{'} 
\nonumber   \\ 
& &{\times}{\langle}\left\{ {\tilde{\rm T}}
{\tilde{\cal {H}}}_{\rm I}(t_1^{+})
{\ldots}{\tilde{\cal {H}}}_{\rm I}(t_n^{+}) \right\}\left\{ {\rm T}
 {\tilde{\cal {H}}}_{\rm I}(t_1^{'-})
{\ldots}{\tilde{\cal {H}}}_{\rm I}(t_m^{'-})
{\tilde{A}(t^{-})}{\tilde{B}(t^{'-})} \right\}{\rangle}_{av}, \nonumber \\
\end{eqnarray}
where ${\langle}{\ldots}{\rangle}_{av}$ 
means ${\rm Tr}[{\tilde{\varrho}}(-\infty){\ldots}]$. 
$\tilde{A}$ denotes an arbitrary operator in the interaction 
 representation.   
 Here, $A(t)$$=S(0,t)\tilde{A}(t)S(t,0)$ and 
${\varrho}_H(0)$$={\tilde{\varrho}}(0)$ $=S(0,-{\infty})
{\tilde{\varrho}}$$(-{\infty})S(-{\infty},0)$.  
In this case, the operators ${\tilde{A}(t^{-})}$ 
and ${\tilde{B}(t^{'-})}$ in Eq.( 2.26 ) both 
are on the $(-)$ branch from $t=-{\infty}$ to $t={\infty}$ 
of the time contour in Fig. 2.1. 

 \ \

Using the last expression in Eq.( 2.26 ), 
the real-time perturbative expansion is 
executed diagrammatically by the help of the Wick's theorem. 
On the diagrammatical perturbative expansion, 
the diagrammatical methods written by Lifshitz and Pitaevskii.[4]
are extended. 

\begin{figure}[ht]
\begin{center}
\includegraphics[width=5.5cm]{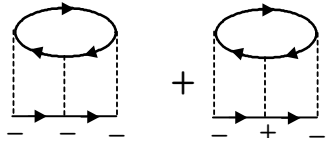}
\caption{The extension of the diagrammatic technique 
of Lifshitz and Pitaevskii. ${\Sigma}^{--(3)}_{ph}$
 brought as the sum over terms going from time $(-)$ to time $(-)$
by way of times $(-)$ and $(+)$. 
}
\end{center}
\end{figure}
\begin{figure}[ht]
\begin{center}
\includegraphics[height=2.5cm,width=3.0cm]{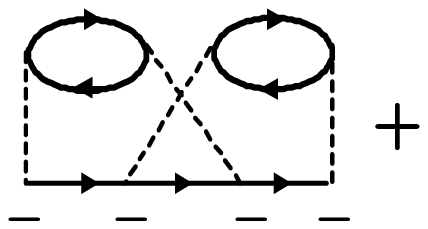}
\includegraphics[height=2.5cm,width=3.0cm]{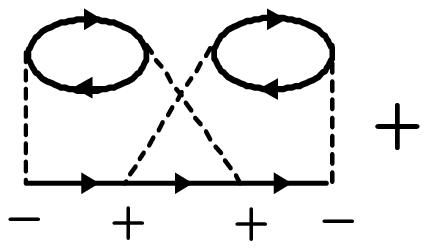}
\includegraphics[height=2.5cm,width=3.0cm]{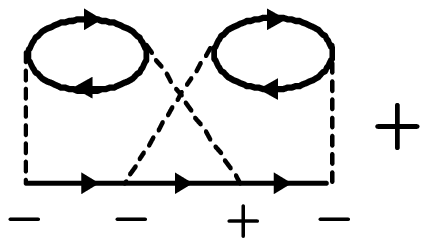}
\includegraphics[height=2.5cm,width=3.0cm]{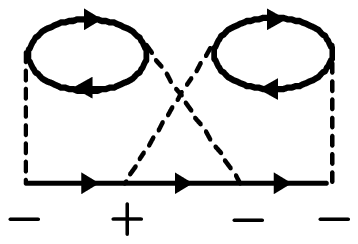}
\caption{${\Sigma}^{--(4)}_{d}$ brought as the sum of terms passing by way of times, 
 $(-)(-),$ $(+)(+),$ $(-)(+)$ and $(+)(-)$.}
\end{center}
\end{figure}

There the summation over terms in all times 
is taken. For example, the third-order self-energy 
${\Sigma}^{--(3)}_{ph}$ is brought  
as the sum over terms going from time $(-)$ to time $(-)$
by way of times $(-)$ and $(+)$ as illustrated in Fig. 2.2.  
 See Fig.2.1 again.  It should be noted that 
the times $(-)$ are on the $(-)$ branch from $t=-{\infty}$ to $t={\infty}$ 
of the time contour in Fig. 2.1 and the times $(+)$ are 
on the $(+)$ branch from $t={\infty}$ to $t=-{\infty}$ 
of the time contour.  
In addition, for the fourth-order self-energy, ${\Sigma}^{--(4)}_{d},$ 
the sum of terms passing through times, 
 $(-)(-),$ $(+)(+),$ $(-)(+)$ and $(+)(-)$ is taken, 
 as shown in Fig. 2.3.

\subsection{Schwinger-Keldysh Formalism } 

 \ \ 

The Dyson's equation for the Keldysh formalism is 
given by 
\begin{eqnarray}
{\bf G}&=&  \ \  {\bf g} \ \  +  \ \ {\bf g}
 \ \ {\bf{\Sigma}} \ \ {\bf G},  
\end{eqnarray}
where 
\begin{eqnarray}
{\bf G}=\left[ \begin{array}
{ll}
G^{--} & G^< \\
 G^> & G^{++} \\
\end{array} 
\right],  \ \
{\bf {\Sigma}}=\left[ \begin{array}
{ll}{\Sigma}^{--} & {\Sigma}^< \\
{\Sigma}^> & {\Sigma}^{++} \\
\end{array} 
    \right].  \nonumber  
\end{eqnarray}

 \ \ 

The nonequilibrium Green's functions are defined by 
\begin{eqnarray}
& & G^{--}(t_1,t_2)\,{\equiv}\,-i{\langle}
{\rm T}\hat{d}(t_1)
 \hat{d}^{\dag}(t_2){\rangle},  \\
& &   G^{++}(t_1,t_2)\,{\equiv}\,
-i{\langle}{\rm \tilde{T}}\hat{d}(t_1)
 \hat{d}^{\dag}(t_2){\rangle},   \\
& &    G^> (t_1,t_2) \,{\equiv}\,
-i{\langle}\hat{d}(t_1)
 \hat{d}^{\dag}(t_2){\rangle},  \\
& &   G^< (t_1,t_2)\,{\equiv}\,i
{\langle}\hat{d}^{\dag}(t_2) \hat{d}(t_1){\rangle}.   
\end{eqnarray}
As shown in Eq. ( 2.26 ), 
the operators in Eq. ( 2.28 ) both are on the $(-)$ 
branch from $t=-{\infty}$ to $t={\infty}$ 
of the time contour in Fig. 2.1. 
In the same way, those in Eq. ( 2.29 ) both are on the $(+)$ 
branch from $t={\infty}$ to $t=-{\infty}$ of the time contour, 
and those in Eq. ( 2.30 ) are on the $(+)$ branch and on the $(-)$ branch, 
respectively. In Eq. ( 2.31 ), they are on the $(-)$ branch 
and on the $(+)$ branch, respectively. 

 \ \ 

Additionally, retarded and advanced Green's functions 
are defined by 
\begin{eqnarray}
 & & G^r (t_1,t_2)\,{\equiv}\,-i{\theta}(t_1-t_2)
 {\langle}\{\hat{d}(t_1),
 \hat{d}^{\dag}(t_2)\}{\rangle},  \\
& &  G^a (t_1,t_2)\,{\equiv}\,i{\theta}(t_2-t_1)
 {\langle}\{\hat{d}(t_1),
 \hat{d}^{\dag}(t_2)\}{\rangle}.  
\end{eqnarray}
Here, the curly brackets signifies anticommutator.
The Dyson's equations for retarded and advanced 
Green's functions are given by 
\begin{eqnarray}
G^r&=&  \ \   g^r \ \  +  \ \  g^r
 \ \ {\Sigma}^r \ \  G^r,  \\
G^a&=&  \ \   g^a \ \  +  \ \  g^a
 \ \ {\Sigma}^a \ \  G^a.  
\end{eqnarray} 
As the necessity to Eqs. ( 2.34 ) and ( 2.35 ),  
the self-energies ${\Sigma}^r$ and ${\Sigma}^a$ 
must be retarded and advanced functions in time, 
respectively.  

 \ \ 

 In accordance with the ordinary procedure of nonequilibrium perturbative 
formalism,[3,4, 34,35,37] for the Dyson's equation for the Keldysh formalism, 
Eq. ( 2.27 ), the transformation is carried out: 
by use of  
\begin{eqnarray}
{\bf L}=[{\bf L^{\dag}}]^{-1}={\frac {1}{\sqrt{2}}}\left[ \begin{array}
{ll}
1 & -1 \\
1 & 1 \\
\end{array} 
\right], \nonumber  
\end{eqnarray}
\begin{eqnarray}
{\bf {G}}=\left[ \begin{array}
{ll}{G}^{--} & {G}^< \\
{G}^> & {G}^{++} \\
\end{array} 
    \right]  \ \ {\longrightarrow} \ \ 
{\bf L}{\bf {G}}{\bf L^{\dag}}=\left[ \begin{array}
{ll}{0} & {G}^a \\
{G}^r & {G}^K \\
\end{array} 
    \right],  \nonumber  
\end{eqnarray}
 and 
\begin{eqnarray}
{\bf {\Sigma}}=\left[ \begin{array}
{ll}{\Sigma}^{--} & {\Sigma}^< \\
{\Sigma}^> & {\Sigma}^{++} \\
\end{array} 
    \right]  \ \ {\longrightarrow} \ \ 
{\bf L}{\bf {\Sigma}}{\bf L^{\dag}}=\left[ \begin{array}
{ll}{\Omega} & {\Sigma}^r \\
{\Sigma}^a & 0 \\
\end{array} 
    \right].  \nonumber  
\end{eqnarray}
From the definition of Green's functions,  
\begin{eqnarray}
{G}^{r}&=&{G}^{--}-{G}^{<}={G}^{>}-{G}^{++}, \nonumber \\
{G}^{a}&=&{G}^{--}-{G}^{>}={G}^{<}-{G}^{++}, \nonumber
\end{eqnarray}
in other words, 
\begin{eqnarray}
{G}^{r}(t)&=&[{G}^{>}(t)-{G}^{<}(t)]{\theta}(t), \nonumber \\
{G}^{a}(t)&=&[{G}^{<}(t)-{G}^{>}(t)]{\theta}(-t). \nonumber
\end{eqnarray}
${G}^K$ is called the Keldysh Green's function. 
For the self-energies part, 
the following relationship is required: 
\begin{eqnarray}
{\Sigma}^{r}(t)&=&{\Sigma}^{--}(t)+{\Sigma}^{<}(t)
=-{\Sigma}^{++}(t)-{\Sigma}^{>}(t), \\
{\Sigma}^{a}(t)&=&{\Sigma}^{--}(t)+{\Sigma}^{>}(t)
=-{\Sigma}^{++}(t)-{\Sigma}^{<}(t), \\ 
{\Omega}(t)&=&{\Sigma}^{--}(t)+{\Sigma}^{++}(t)
=-{\Sigma}^<(t)-{\Sigma}^>(t).
\end{eqnarray}
The above is  known in general. 
Here, it is uncertain whether or not the requirements of self-energies 
as functions in time are fulfilled. It is because  
the definition of the self-energies in matrix form is not given, 
 as mentioned earlier. 

 \ \ 

 \ \ 

The solution is given from the present work.[43] 
That is explained as follows: 
in the right-hand sides of Eqs. ( 2.36 ), ( 2.37 ) and ( 2.38 ), 
the functions, ${\Sigma}^{--}(t),$ ${\Sigma}^{<}(t),$
${\Sigma}^{++}(t)$ and ${\Sigma}^{>}(t)$ are  
deduced from perturbative expansion with the Wick's theorem. 
At this time, the directions in time are not defined in the 
functions ${\Sigma}^{--}(t),$ ${\Sigma}^{<}(t),$ 
${\Sigma}^{++}(t)$ and ${\Sigma}^{>}(t)$.   
This is due to the following functions  
in dependence upon time:   
\begin{eqnarray}
g^{--}(t)={\theta}(t)g^>(t)+{\theta}(-t)g^<(t), \\
g^{++}(t)={\theta}(t)g^<(t)+{\theta}(-t)g^>(t). 
\end{eqnarray}
For this reason, for the right-hand sides of 
Eqs. ( 2.36 ), ( 2.37 )  and ( 2.38 ),   
the directions in time must necessarily be taken into 
consideration. 

 \ \ 

As the main point, 
the terms in the right-hand sides are taken 
 for sum of retarded and advanced terms:
\begin{eqnarray}
{\Sigma}^{r}(t)&=&[{\Sigma}^{--}(t)+{\Sigma}^{<}(t)]
{\theta}(t)+[{\Sigma}^{--}(t)+{\Sigma}^{<}(t)]{\theta}(-t) \nonumber \\
&=&-[{\Sigma}^{++}(t)+{\Sigma}^{>}(t)]{\theta}(t)
-[{\Sigma}^{++}(t)+{\Sigma}^{>}(t)]{\theta}(-t), 
\end{eqnarray}
\begin{eqnarray}
{\Sigma}^{a}(t)&=&[{\Sigma}^{--}(t)+{\Sigma}^{>}(t)]{\theta}(t)
+[{\Sigma}^{--}(t)+{\Sigma}^{>}(t)]{\theta}(-t) \nonumber \\
&=&-[{\Sigma}^{++}(t)+{\Sigma}^{<}(t)]{\theta}(t)
-[{\Sigma}^{++}(t)+{\Sigma}^{<}(t)]{\theta}(-t),   
\end{eqnarray}
\begin{eqnarray}
{\Omega}(t)&=&[{\Sigma}^{--}(t)+{\Sigma}^{++}(t)]{\theta}(t)+
[{\Sigma}^{--}(t)+{\Sigma}^{++}(t)]{\theta}(-t) \nonumber  \\
&=&-[{\Sigma}^<(t)+{\Sigma}^>(t)]{\theta}(t)
-[{\Sigma}^<(t)+{\Sigma}^>(t)]{\theta}(-t).  
\end{eqnarray}

 \ \ 

For self-energy functions derived by the present 
perturbative expansion via the Wick's theorem, the following 
relations are found: 
\begin{eqnarray}
{\Sigma}^{--}(t){\theta}(t)=-{\Sigma}^{>}(t){\theta}(t),  \\       
{\Sigma}^{++}(t){\theta}(t)=-{\Sigma}^{<}(t){\theta}(t), \\  
{\Sigma}^{--}(t){\theta}(-t)=-{\Sigma}^{<}(t){\theta}(-t),  \\ 
{\Sigma}^{++}(t){\theta}(-t)=-{\Sigma}^{>}(t){\theta}(-t);  
\end{eqnarray} 
these relations have never been known in general.   
When these are substituted into Eqs. ( 2.41 ), ( 2.42 ) and ( 2.43 ), 
then, the advanced term of Eq. ( 2.41 ) and the retarded term of 
Eq. ( 2.42 ) are canceled:  
\begin{eqnarray}
& & [{\Sigma}^{--}(t)+{\Sigma}^{<}(t)]{\theta}(-t)= 
-[{\Sigma}^{++}(t)+{\Sigma}^{>}(t)]{\theta}(-t)=0,  \\
& & [{\Sigma}^{--}(t)+{\Sigma}^{>}(t)]{\theta}(t)=
-[{\Sigma}^{++}(t)+{\Sigma}^{<}(t)]{\theta}(t)=0,  
\end{eqnarray}
so that Equations ( 2.41 ) and ( 2.42 ) reduce to 
\begin{eqnarray}
{\Sigma}^{r}(t)&=&[{\Sigma}^{--}(t)+{\Sigma}^{<}(t)]{\theta}(t)
=-[{\Sigma}^{++}(t)+{\Sigma}^{>}(t)]{\theta}(t), \\
{\Sigma}^{a}(t)&=&[{\Sigma}^{--}(t)+{\Sigma}^{>}(t)]{\theta}(-t)
=-[{\Sigma}^{++}(t)+{\Sigma}^{<}(t)]{\theta}(-t). 
\end{eqnarray}
The retarded and advanced self-energies are 
acquired  as retarded and advanced functions in time, 
respectively; they are the requirements. 
In addition, Equation ( 2.43 ) is certainly reproduced.    

\ \ 

Then, it leads to     
\begin{eqnarray}
{\Sigma}^{r}(t)&=&[{\Sigma}^{<}(t)-{\Sigma}^{>}(t)]{\theta}(t), 
 \\ 
{\Sigma}^{a}(t)&=&[{\Sigma}^{>}(t)-{\Sigma}^{<}(t)]{\theta}(-t); 
\end{eqnarray}
they consist with the expressions in the review by Rammer and Smith.[44] 
The formalism in the review of Rammer and Smith is different from 
  the nonequilibrium perturbative expansion,  
 so that the present method is confirmed to connect with the other 
 formalism.  
By performing the Fourier  transformation for  
Eqs. ( 2.52 ) and ( 2.53 ),  we have 
\begin{eqnarray}
{\Sigma}^{r}(E)-{\Sigma}^{a}(E)={\Sigma}^{<}(E)-{\Sigma}^{>}(E);  
\end{eqnarray}
it proves that the relation stands.  
Since Equation ( 2.54 ) is widely known,   
Equations ( 2.52 ) and ( 2.53 ) 
should generally hold as functions in time.

Besides, 
\begin{eqnarray}
G^<&=&(1+G^r{\Sigma}^{r})g^<(1+G^a{\Sigma}^{a})-
G^r{\Sigma}^{<}G^a,   \\
G^>&=&(1+G^r{\Sigma}^{r})g^>(1+G^a{\Sigma}^{a})-
G^r{\Sigma}^{>}G^a     
\end{eqnarray}
and 
\begin{eqnarray}
G^K&=&(1+G^r{\Sigma}^{r})g^K(1+G^a{\Sigma}^{a})+
G^r{\Omega}G^a  
\end{eqnarray}
still work.  

 \ \ 

As mentioned above, the present solution is 
in accordance with the generally known relations.
It indicates that the present solution has validity.

\chapter{Expressions of Self-Energy for Anderson model }
\section{Anderson model} 

 \ \ 

 \ \ 

We consider equilibrium and nonequilibrium stationary states.  
Nonequilibrium state is caused by finite bias voltage, 
that is, the difference of chemical potentials;    
after bias voltage was turned on, long time has passed enough 
to reach stationary states. 
Since the states are stationary, the Hamiltonian has no 
time dependence. The system is  described by 
the Anderson model linking to leads. 
The impurity ( the quantum dot ) with on-site energy $E_0$ 
and the Coulomb interaction $U$ 
is connected to the left and right leads   
by the mixing matrix elements, $v_L$ and $v_R$.  
The system is illustrated below. 

 \ \ 

 \ \ \ \ 
\begin{tabular}{c}
 U=0 \\
\fbox{Left Lead}
\end{tabular} 
\underline{ \ \ \ \ $v_L$ \ \ \ \  } 
\begin{tabular}{c}
 U${\neq}$0 \\
\Ovalbox{
\begin{tabular}{c}
Quantum \\Dot
\end{tabular}}
\end{tabular} 
\underline{ \ \ \ \ $v_R$ \ \ \ \  } 
\begin{tabular}{c}
 U=0 \\
\fbox{Right Lead}
\end{tabular} 

 \ \ 

 \ \ 

 \ \ 

 The Anderson Hamiltonian is given by 
\begin{eqnarray}
{\cal H}=& & E_0 \sum_{\sigma}\hat{n}_{d{\sigma}}
+ {\mu}_L \sum_{\sigma}\hat{n}_{L{\sigma}}
+ {\mu}_R \sum_{\sigma}\hat{n}_{R{\sigma}}
+U \bigl(\hat{n}_{d\uparrow}
-{\langle}\hat{n}_{d\uparrow}{\rangle}\bigr)
\bigl(\hat{n}_{d\downarrow}-{\langle}\hat{n}_{d\downarrow}{\rangle}\bigr)
\nonumber  \\
  & & -\sum_{{\sigma}}v_{L}\bigl(
  \hat{d}_{{\sigma}}^{\dag}\hat{c}_{L{\sigma}}+{\rm H.c.}\bigr) 
-\sum_{{\sigma}}v_{R}\bigl(\hat{d}_{{\sigma}}^{\dag}\hat{c}_{R{\sigma}}
+{\rm H.c.}\bigr). 
\end{eqnarray}
$\hat{d}^{\dag}$ ($\hat{d}$) is creation (annihilation) operator 
 for electron on the impurity, and 
$\hat{c}_{L}^{\dag}$ and $\hat{c}_{R}^{\dag}$ 
($\hat{c}_{L}$ and $\hat{c}_{R}$) are creation (annihilation) operators 
in the left and  right leads, respectively. 
${\sigma}$ is index for spin.  
The chemical potentials in the isolated left and right leads  
are ${\mu}_{L}$ and ${\mu}_{R},$ respectively. 
  The applied voltage is, therefore 
 defined by $eV\, {\equiv}\, {\mu}_L-{\mu}_R$.

 \ \

We consider that the band-width of left and right leads 
is large infinitely, so that the coupling functions, 
 ${\Gamma}_L$ and ${\Gamma}_R$ can be taken to be independent of energy, $E$.
On-site energy $E_0$ is set being canceled with 
the Hartree term, {\it i.e.} the first-order contribution to self-energy for  
electron correlation, as mentioned later.  

Accordingly, the Fourier components of the noninteracting 
 ( unperturbed ) Green's functions reduce to  
\begin{eqnarray}
 g^r(E)&=&{\frac{1}{E+i{\Gamma},}} \\
 g^a(E)&=&{\frac{1}{E-i{\Gamma}.}} 
\end{eqnarray}
where ${\Gamma}=( {\Gamma}_L+{\Gamma}_R )/ 2$.
Hence, the inverse Fourier components  
 can be written by 
\begin{eqnarray}
g^r(t)=-i{\theta}(t)e^{-{\Gamma}t}, \\ 
g^a(t)=i{\theta}(-t)e^{{\Gamma}t}.   
\end{eqnarray}
In addition, by solving the Dyson's equation Eq. ( 2.27  ), we have
\begin{eqnarray}
 & & g^<(E)=g^r(E)\, 
\bigl[\, if_L(E){\Gamma}_L+if_R(E){\Gamma}_R \, \bigr]\, g^a(E), \\
 & & g^>(E)=g^r(E)\, 
\bigl[\, i(f_L(E)-1){\Gamma}_L+i(f_R(E)-1){\Gamma}_R \, \bigr]\, g^a(E). 
\end{eqnarray}
$f_L$ and $f_R$ are the Fermi distribution functions 
in the isolated left and right leads, respectively. 
By Eqs. ( 3.6 ) and ( 3.7 ), the nonequilibrium state is  
introduced effectively as the superposition of the left and right leads.  
In this case, the effective Fermi distribution function can be 
expressed by[8]
\begin{eqnarray}
f_{\rm eff}(E)={\frac {f_L(E){\Gamma}_L+f_R(E){\Gamma}_R}
{{\Gamma}_L+{\Gamma}_R}}.
\end{eqnarray}
This effective Fermi distribution function is reasonable 
because it is considered that leads have the continuous energy states  
and the two chemical potentials are not two localized states.

\section{Self-Energy } 

 \ \ 

The expressions for self-energies of the Coulomb 
interaction of the Anderson model are formulated 
by the method of the nonequilibrium perturbation 
 theory based on the adiabatic theorem, 
explained in Chapter 2. 
Practically, the perturbative expansion 
is done with respect to the Coulomb interaction term 
of Eq. ( 3.1 ) using the last expression of Eq. ( 2.26 )
in view of the Dyson's equation Eq. ( 2.27 ) and each diagram.  
In ${\Sigma}^{--}(t),$ ${\Sigma}^{<}(t),$
${\Sigma}^{++}(t)$ and ${\Sigma}^{>}(t)$ obtained 
by the present solution,  Equations (2.44)-(2.47) are satisfied. 
Every formulated ${\Sigma}^{r}(E)$ and ${\Sigma}^{a}(E)$ 
are in the relation of the complex conjugate each other. 

\subsection{First-Order Contribution }

 \ \ 

For the Anderson model, 
the first-order contribution, the Hartree term  
can be written by  
\begin{eqnarray}
{\Sigma}^{r(1)}(E)={\Sigma}^{a(1)}(E)
&=&U{\langle}n{\rangle}=
U\int {\frac{dE}{2{\pi}i}}G^{<}(E), 
\end{eqnarray}
where ${\langle}n{\rangle}$ is charge density.  

\subsection{Second-Order Contribution }

 \ \ 

The second-order self-energies are expressed by 
\begin{eqnarray}
{\Sigma}^{r(2)}(E) =
U^2\int^{\infty}_{0}{dt_{1}}    
e^{iEt_{1}}\left[ 
\begin{array}
{llll}
\ \ g^{>}(t_{1})g^>(t_{1}) g^<(-t_{1}) \\ 
- g^<(t_{1})g^<(t_{1}) g^>(-t_{1})   
\end{array} \right] \nonumber \\ 
=U^2\int^{\infty}_{0}{dt_{1}}    
e^{iEt_{1}}\left[ 
\begin{array}
{llll}
\ \ g^{\pm}(t_{1})g^>(t_{1}) g^<(-t_{1}) \\ 
+ g^<(t_{1})g^{\pm}(t_{1}) g^>(-t_{1})   \\
+ g^<(t_{1})g^>(t_{1}) g^{\pm}(-t_{1})
\end{array} \right],  
\end{eqnarray}
\begin{eqnarray}
{\Sigma}^{a(2)}(E) =
U^2\int^{0}_{-\infty}{dt_{1}}    
e^{iEt_{1}}\left[ 
\begin{array}
{llll}
\ \ g^<(t_{1})g^{<}(t_{1}) g^>(-t_{1})   \\
- g^>(t_{1})g^>(t_{1}) g^{<}(-t_{1})
\end{array} \right] \nonumber \\  
=U^2\int^{0}_{-\infty}{dt_{1}}    
e^{iEt_{1}}\left[ 
\begin{array}
{llll}
\ \ g^{\pm}(t_{1})g^>(t_{1}) g^<(-t_{1}) \\ 
+ g^<(t_{1})g^{\pm}(t_{1}) g^>(-t_{1})   \\
+ g^<(t_{1})g^>(t_{1}) g^{\pm}(-t_{1})
\end{array} \right].  
\end{eqnarray}
Here $g^{\pm}(t)=g^r(t)+g^a(t),$ that is, 
$g^{+}(t)=g^r(t)=-i{\theta}(t)e^{-{\Gamma}t}$ 
for $t{\ge}0$ and $g^{-}(t)=g^a(t)=i{\theta}(-t)e^{{\Gamma}t}$ 
for $t<0$. Additionally, $g^<(t)$ and $g^>(t)$
 are the inverse Fourier components of Eqs. ( 3.6 ) and ( 3.7 ). 
Figure 3.1 shows the diagram for the second-order self-energy. 
These expressions for equilibrium agree exactly with those   
deduced from the  Matsubara imaginary-time perturbative expansion  
for equilibrium and analytical continuity by 
Zlati{\'c} {\it et al.}[45].  
As shown numerically later, the second-order contribution
 coincide with those brought out by Hershfield {\it et al.}[8]. 

 \ \ 

In the symmetric equilibrium case,
the asymptotic behavior at low energy is expressed by 
\begin{eqnarray}
 {\Sigma}^{r(2)}(E){\simeq}
-{\Gamma}\left(3- {\frac{{\pi}^2}{4}}\right)
\left(\frac{U}{{\pi}{\Gamma}}\right)^2 \frac{E}{\Gamma}
-i\frac{\Gamma}{2}\left(\frac{U}{{\pi}{\Gamma}}\right)^2
\left(\frac{E}{\Gamma}\right)^2,
\end{eqnarray}
the exact results based on the Bethe ansatz method.[46,47] 
\begin{figure}[htbp]
\begin{center}
\includegraphics[width=5.0cm]{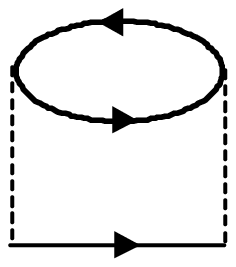}
\caption{The second-order self-energy }
\end{center}
\end{figure}

\subsection{Third-Order and Fourth-Order Contributions }

 \ \ 

There are two kinds of the third-order 
contributions as illustrated in Fig. 3.2. 
\begin{eqnarray}
 {\Sigma}^{r(3)}_{pp}(E)
&=&U^3\int^{\infty}_{0} {dt_{1}}\int^{\infty}_{-\infty}
{dt_{2}}e^{iEt_1}\, 
\left[ 
\begin{array}
{llll}
 g^<(-t_1) g^>(t_1-t_2)
 g^>(t_1-t_2) \nonumber \\ 
 -  g^>(-t_1) g^<(t_1-t_2)
 g^<(t_1-t_2) \nonumber \\ 
\end{array} 
\right] \\
& & {\times} 
\left[ 
\begin{array}
{llll}
 g^{\pm}(t_2) g^>(t_2)  
 + g^<(t_2) g^{\pm}(t_2)  
\end{array} 
\right], 
\end{eqnarray}
\begin{eqnarray}
 {\Sigma}^{a(3)}_{pp}(E)
&=&U^3\int^{0}_{-\infty} {dt_{1}}\int^{\infty}_{-\infty}
{dt_{2}}e^{iEt_1}\, 
\left[ 
\begin{array}
{llll}
g^>(-t_1) g^<(t_1-t_2)
 g^<(t_1-t_2) \nonumber \\
- g^<(-t_1) g^>(t_1-t_2)
 g^>(t_1-t_2)  
\nonumber \\ 
\end{array} 
\right] \\
& & {\times} 
\left[ 
\begin{array}
{llll}
 g^{\pm}(t_2) g^>(t_2)  
 + g^<(t_2) g^{\pm}(t_2)  
\end{array} 
\right]. 
\end{eqnarray}
\begin{eqnarray}
 {\Sigma}^{r(3)}_{ph}(E)
&=&U^3\int^{\infty}_{0} {dt_{1}}\int^{\infty}_{-\infty}
{dt_{2}}e^{iEt_1}\, 
\left[ 
\begin{array}
{llll}
 g^>(t_1) g^>(t_1-t_2)
 g^<(t_2-t_1) \nonumber \\ 
 -  g^<(t_1) g^<(t_1-t_2)
 g^>(t_2-t_1) \nonumber \\ 
\end{array} 
\right] \\
& &{\times} 
\left[ \begin{array}
{llll}
 g^{\pm}(t_2) g^<(-t_2)  
 + g^<(t_2) g^{\pm}(-t_2)  
\end{array} 
\right], 
\end{eqnarray}
\begin{eqnarray}
 {\Sigma}^{a(3)}_{ph}(E)
&=&U^3\int^{0}_{-\infty} {dt_{1}}\int^{\infty}_{-\infty}
{dt_{2}}e^{iEt_1}\, 
\left[ 
\begin{array}
{llll}
 g^<(t_1) g^<(t_1-t_2)
 g^>(t_2-t_1) \nonumber \\ 
- g^>(t_1) g^>(t_1-t_2)
 g^<(t_2-t_1) \nonumber \\ 
\end{array} 
\right] \\
& &{\times} 
\left[ \begin{array}
{llll}
 g^{\pm}(t_2) g^<(-t_2)  
 + g^<(t_2) g^{\pm}(-t_2)  
\end{array} 
\right]. 
\end{eqnarray}

 \ \ 

Equations (3.13)-(3.16)  for equilibrium state agree exactly with those   
derived from the  Matsubara imaginary-time perturbative expansion  
for equilibrium and analytical continuity by 
Zlati{\'c} {\it et al.}[45].  
As mentioned later, it is numerically confirmed that 
the third-order contribution is canceled for the symmetric Anderson model;  
this is compatible with both the results deduced from the 
Yamada-Yosida theory[30,47,48] and gained  
on the basis of the Bethe ansatz method[46].  

\begin{figure}[htbp]
\begin{center}
\includegraphics[width=6.0cm]{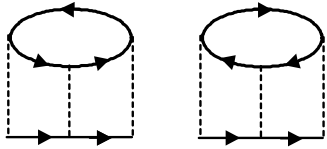}
\caption{Two kinds of the third-order self-energies: 
${\Sigma}^{(3)}_{pp}$(Left) and ${\Sigma}^{(3)}_{ph}$}
\end{center}
\end{figure}

 \ \ 

Furthermore, the fourth-order contribution to the 
self-energy is formulated. ( See Appendix. )
The twelve terms for the proper fourth-order self-energy 
 can be divided into four groups, each of which comprises   
 three terms, corresponding diagrams 
 Fig. 3.3 (a)-(c),  (d)-(f), (g)-(i) and (j)-(l), 
respectively. 

\begin{figure}[htbp]
\begin{center}
\includegraphics[width=7.0cm]{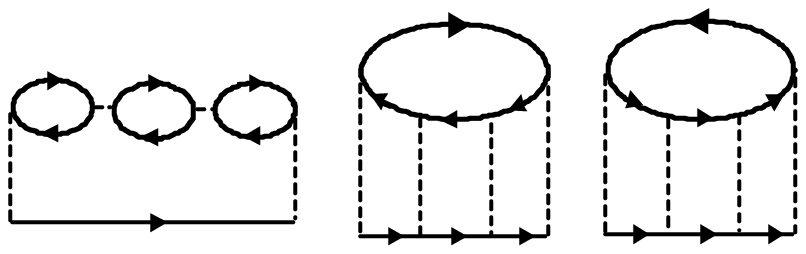}
\includegraphics[width=7.0cm]{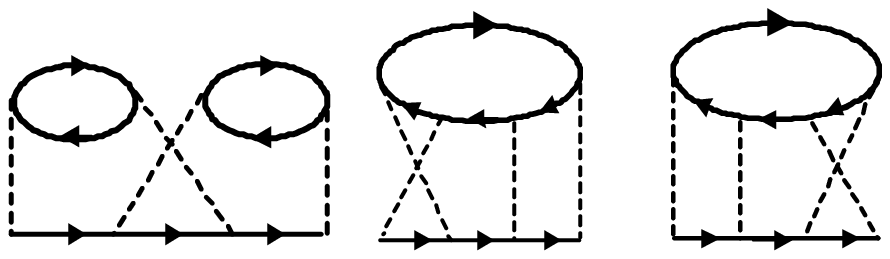}
\end{center}
\begin{center}
\includegraphics[width=7.0cm]{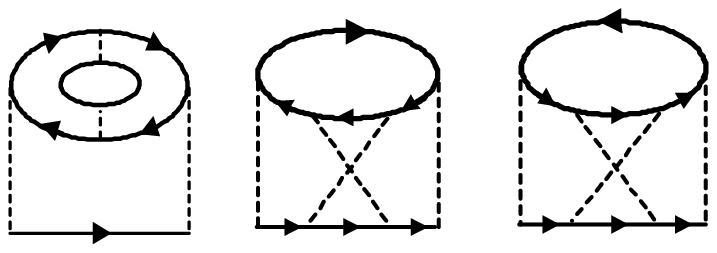}
\includegraphics[width=7.0cm]{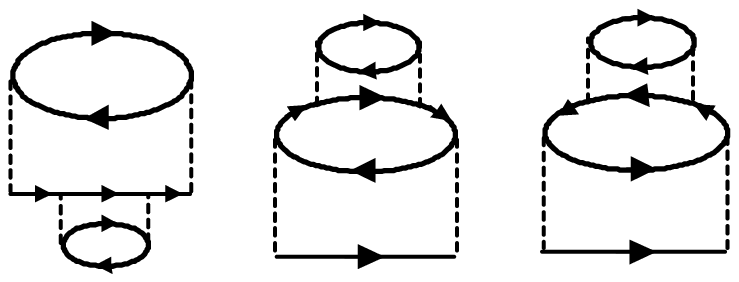}
\caption{The twelve terms for the proper fourth-order self-energy 
 divided into four groups: (a)-(c),  (d)-(f), (g)-(i), 
and (j)-(l).}
\end{center}
\end{figure}

 \ \ 

 \ \ 

 \ \ 

 \ \ 

 \ \ 

For symmetric Anderson model at equilibrium, 
the asymptotic behavior at low energy  
is approximately in agreement with those based on 
the Bethe ansatz method[46]:
\begin{eqnarray}
 {\Sigma}^{r(4)}(E){\simeq}
-{\Gamma}\left(105- {\frac{45{\pi}^2}{4}}
+{\frac{{\pi}^4}{16}}\right)
\left(\frac{U}{{\pi}{\Gamma}}\right)^4 \frac{E}{\Gamma}
-i\frac{\Gamma}{2}\left(30-3{\pi}^2\right)\left(\frac{U}
{{\pi}{\Gamma}}\right)^4
\left(\frac{E}{\Gamma}\right)^2.  \nonumber \\
\end{eqnarray}

The fourth-order contribution has not 
been clarified well.   
In particular, the behavior for nonequilibrium 
 state has almost been unknown.  In Chapter 4, 
the numerical results are shown and discussed.

\chapter{Numerical Results and Discussion }
\section{Self-Energy  }

 \ \ 

The third-order terms, Eqs. ( 3.13 )-( 3.16 ) are canceled 
under electron-hole symmetry  
 not only at equilibrium but also at nonequilibrium: 
\begin{eqnarray}
{\Sigma}^{r(3)}_{ph}(E)&=&-{\Sigma}^{r(3)}_{pp}(E), \nonumber \\  
{\Sigma}^{a(3)}_{ph}(E)&=&-{\Sigma}^{a(3)}_{pp}(E). \nonumber  
\end{eqnarray}
As a consequence, the third-order contribution to self-energy 
vanishes in the symmetric case.  
It is consistent with the results of Refs. [30, 47, 48] based on the 
Yamada-Yosida theory that all odd-order contributions 
except the Hartree term become null for equilibrium
 in the symmetric single-impurity Anderson model;  
probably, it is just the same with nonequilibrium state.
 On the other hand, the third-order terms contribute to 
the asymmetric system where electron-hole symmetry breaks  
and furthermore, the third-order terms for spin-up 
and for spin-down contribute respectively when 
the spin degeneracy is lifted for example, by  magnetic field. 
\begin{figure}[ht]
\includegraphics[width=7.0cm]{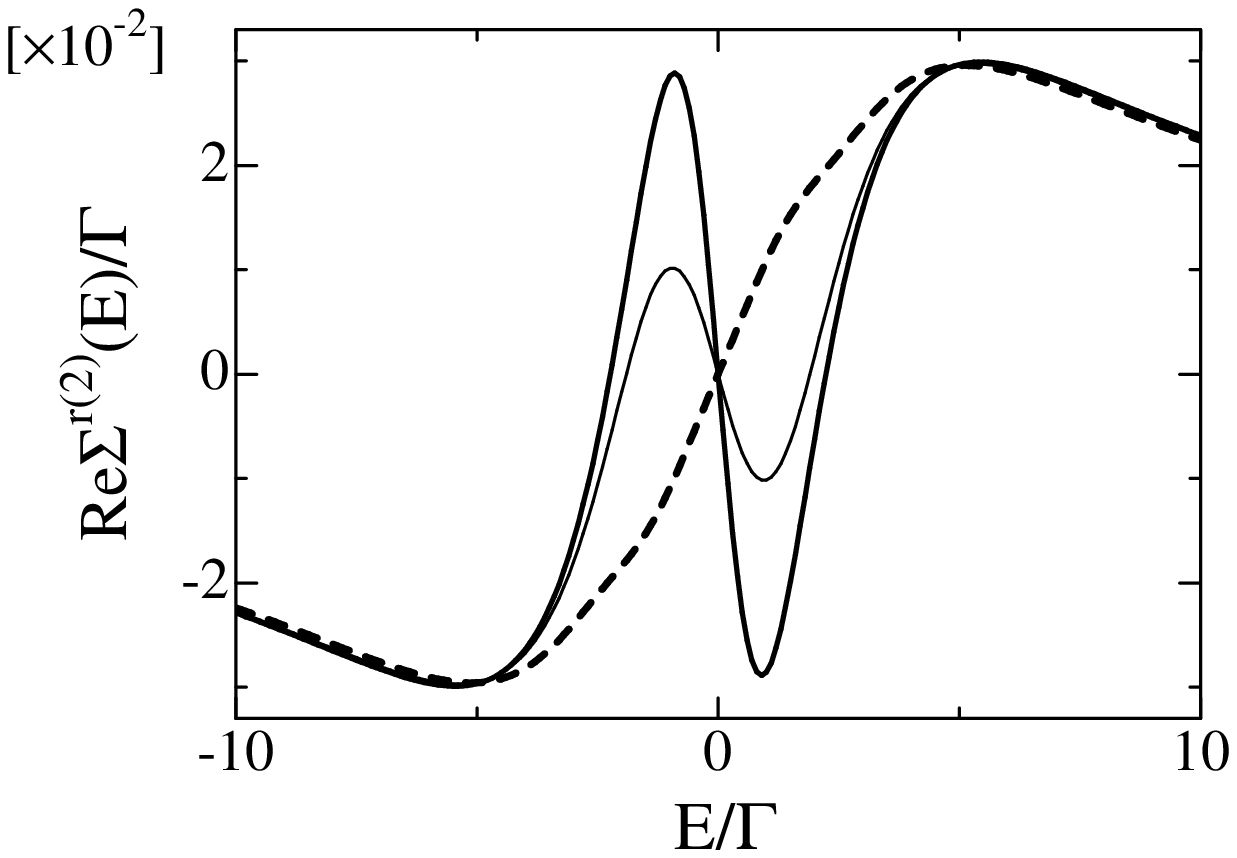}
\includegraphics[width=7.0cm]{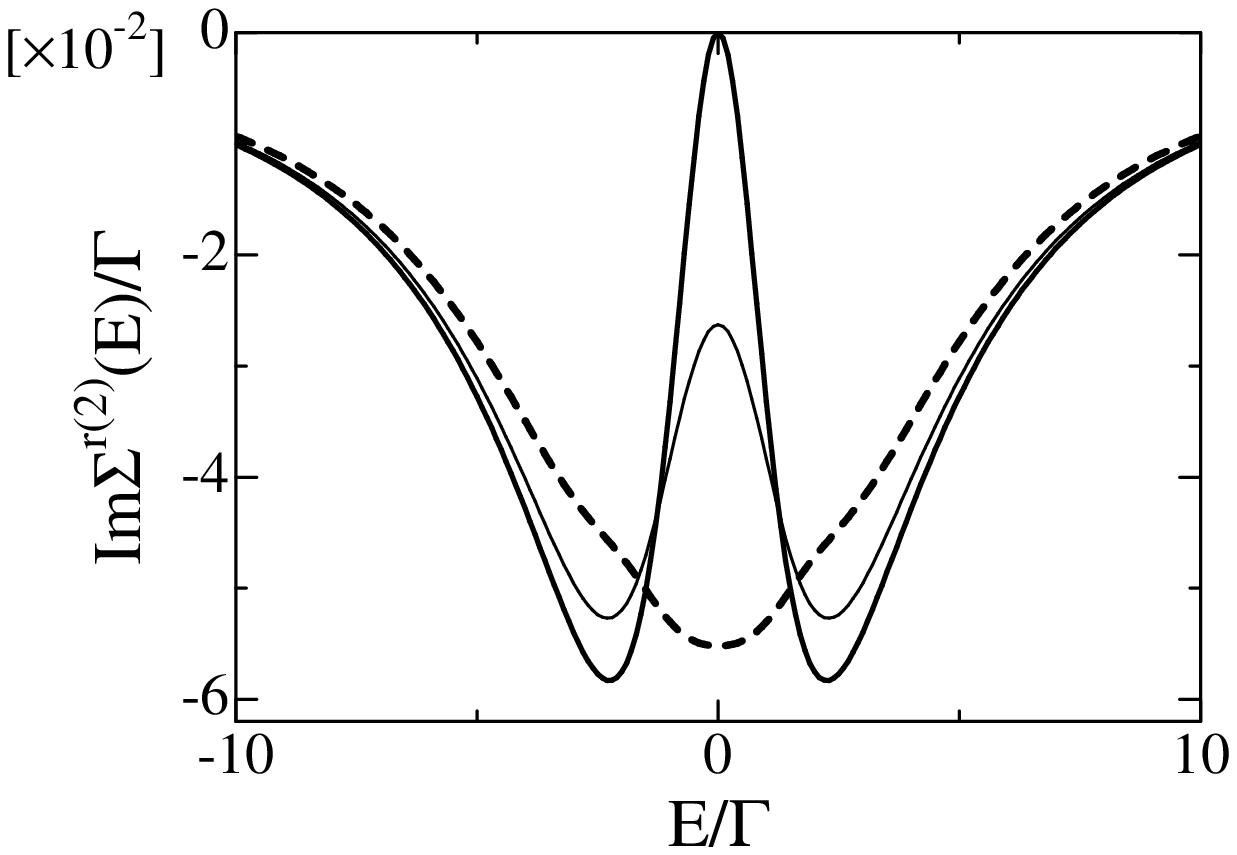}
\caption{
The second-order self-energy for the symmetric Anderson model 
at $U/{\Gamma}=1.0$ and zero temperature. 
Left:real part and Right:imaginary part  
 at equilibrium ( solid line ), 
$eV/{\Gamma}=1.0$ ( thin solid line ) and 
$eV/{\Gamma}=2.0$ ( dashed line ). 
}
\end{figure}
\begin{figure}[ht]
\includegraphics[width=7.0cm]{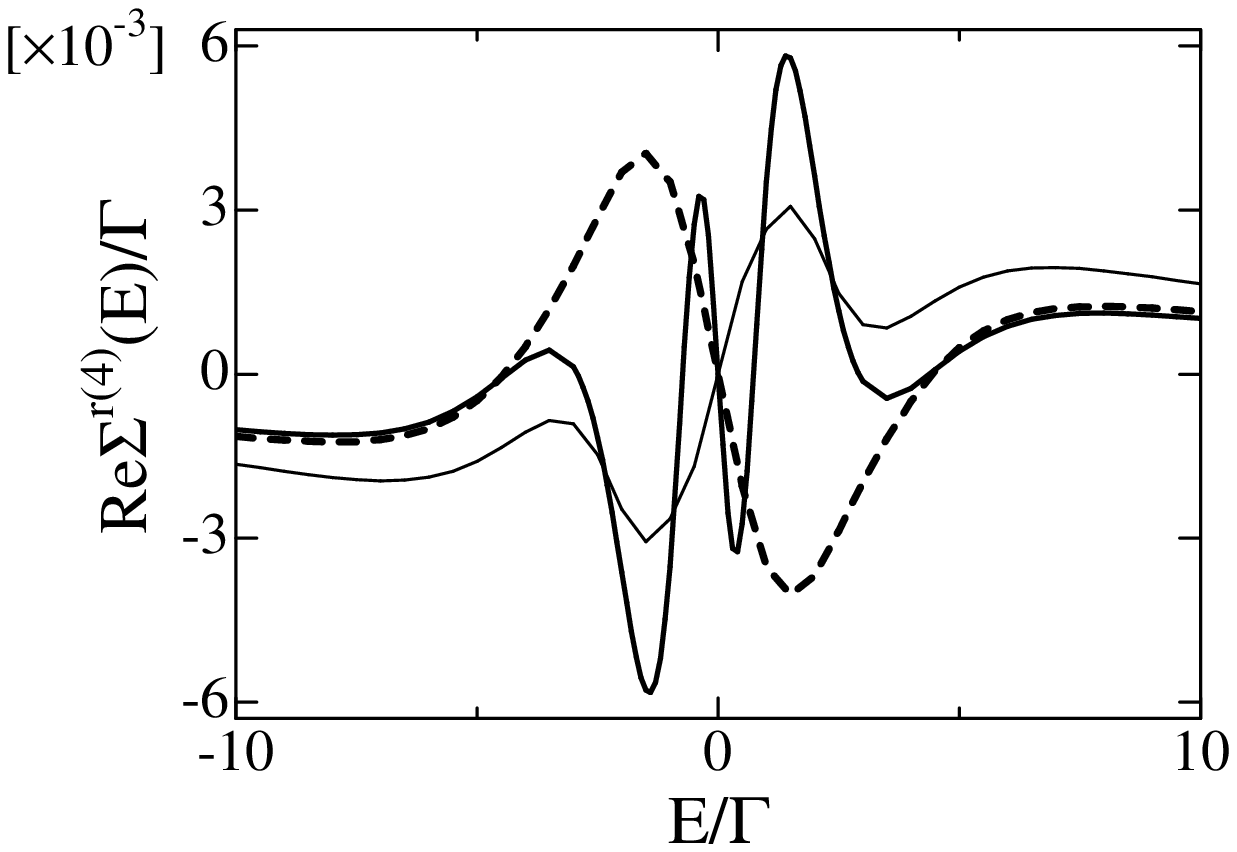}
\includegraphics[width=7.0cm]{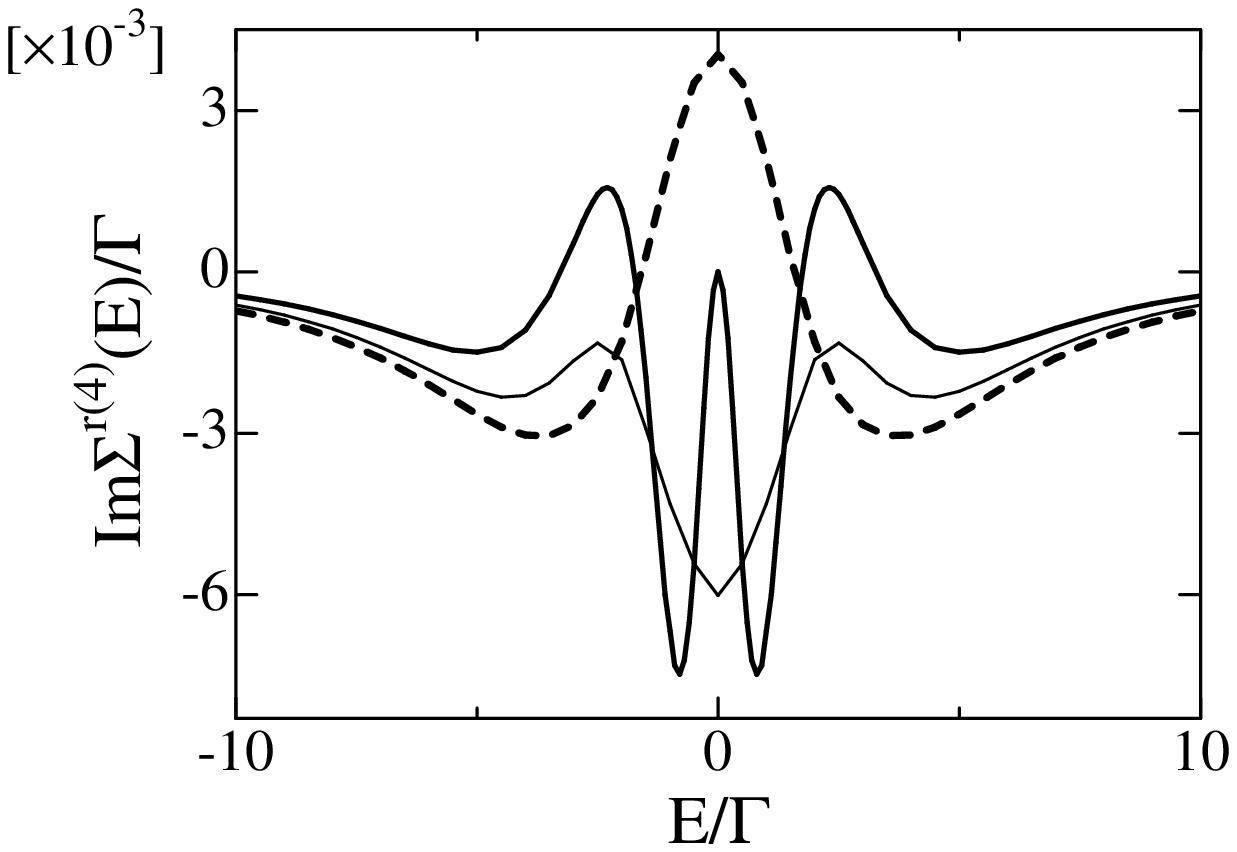}
\caption{
The fourth-order self-energy for the symmetric Anderson model 
at $U/{\Gamma}=1.0$ and zero temperature. 
Left:real part and Right:imaginary part 
at equilibrium ( solid line ), 
$eV/{\Gamma}=1.0$ ( thin solid line ), and 
$eV/{\Gamma}=2.0$ ( dashed line ). 
The fourth-order contribution for equilibrium 
has the same but narrow curves at low energy with those of the second-order 
contribution.
}
\end{figure}
For the fourth-order contribution,  
three terms which constitute each of four groups 
contribute equivalently under electron-hole symmetry. 
Moreover,  to the asymmetric system, 
the terms brought by the diagrams of Fig. 3.3(a) and (b) 
contribute equivalently  and 
the terms by the diagrams of Fig. 3.3(j) and (k)
make equivalent contribution, and  
the rest, the eight terms contribute respectively. 
Further, the twenty-four terms for spin-up and spin-down 
take effect severally in the presence of magnetic field. 

 The second-order and the fourth-order contributions to self-energy 
for zero temperature symmetric Anderson model 
are plotted in Fig. 4.1  and in Fig. 4.2, respectively. 
Equation ( 3.12  ) represents  the curves around $E=0$ 
  denoted by solid line in Fig. 4.1, 
and Equation ( 3.17 ) represents approximately 
those  shown in Fig. 4.2. 
The curves of the second-order self-energy shown  
 in Fig. 4.1 are identical with those of expressions 
derived by Hershfield {\it et al.}[8]. 
In comparison of Fig. 4.2 with Fig. 4.1, 
 it is found that the fourth-order contribution for equilibrium 
has the same but narrow curves at low energy with those of the second-order 
contribution.  In addition, the broad curves are attached 
at high energy for the fourth-order self-energy. 
( The higher contribution is, the more should the curves 
oscillate as a function of energy. ) 
When the voltage, $eV/{\Gamma}$ exceeds ${\sim}2.0,$ 
the behavior of curves of self-energy changes distinctly and  
 comes to present a striking contrast to that for the second-order 
contribution. Especially, the curve for the imaginary part 
of the fourth-order contribution  rises up  
with maximum at $E=0$.   
On the other hand, for the second-order contribution, 
 a valley appears with minimum at energy of zero$-$it  
 is quite the contrary. Moreover, 
from these results, 
it is expected that the sixth-order contribution 
to imaginary part of self-energy has minimum at $E=0$. 
By reason of these, the perturbative expansion is 
hard to converge for $eV/{\Gamma}>{\sim}2.0,$ as 
mentioned later.

\section{Current Conservation }

 \ \ 

In this section, the problem on the current conservation is 
described below. 
In Ref. [8], it is shown that the continuity of current 
entering and leaving the impurity  
 stands exactly at any strength of $U$   
within the approximation up to the second-order 
for the symmetric single-impurity Anderson model.  
In comparison of Fig. 4.2 with Fig. 4.1, 
it is found that curves of fourth-order 
self-energy have the symmetry similar to those of 
the second-order. 
From this, it is anticipated that the current conservation are 
kept perfectly with approximation up to the 
fourth-order in the single-impurity system where 
electron-hole symmetry holds. 
The continuity of current can be maintained   
perfectly in single-impurity system as far as electron-hole 
symmetry stands. On the other hand, current comes to fail 
to be conserved with increasing $U$ 
in asymmetric single-impurity case and 
in two-impurity case. 

 \ \ 

\section{Spectral Function  }
\subsection{For Second-Order Self-Energy  }

 \ \ 

\begin{figure}[ht]
\begin{center}
\includegraphics[width=8.0cm]{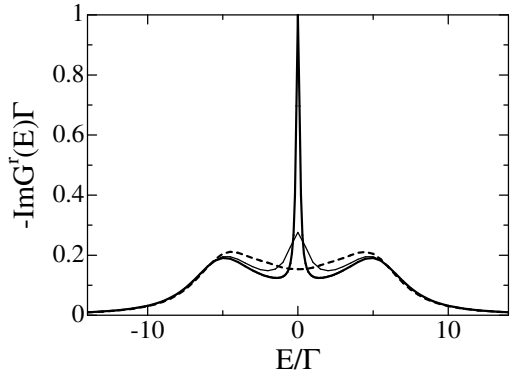}
\caption{
The spectral function with the second-order self-energy  
at  $U/{\Gamma}=10.0$ for the symmetric Anderson model  
 at equilibrium ( solid line ), 
$eV/{\Gamma}=1.0$ ( thin solid line ) and 
$eV/{\Gamma}=2.0$ ( dashed line ). 
 }
\end{center}
\end{figure}
The spectral function with the second-order self-energy 
is widely known. [8][47] 
It is  plotted for $U/{\Gamma}=10.0$ and zero temperature in Fig. 4.3.  
For equilibrium, the Kondo peak at energy of zero is very sharp 
and the two-side broad peaks appear at $E {\simeq}$ ${\pm}U/2$. 
The curve for $eV=0$ is identical with that shown in Ref. [47]. 
As $eV$ becomes higher than the Kondo temperatures, 
 $k_BT_K$[49], 
the Kondo peak becomes lower and is lost finally, 
while the two-side broad peaks rise at $E{\simeq}$ 
${\pm}U/2$.[8]

\subsection{For Self-Energy up to Fourth-Order for Equilibrium}

 \ \ 

\begin{figure}[ht]
\begin{center}
\includegraphics[width=8.0cm]{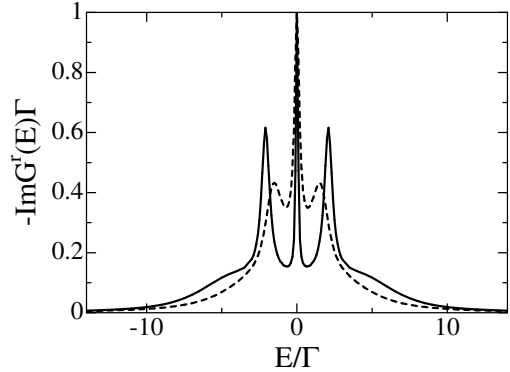}
\caption{
The spectral function with  self-energy  up to the fourth-order
at equilibrium for the symmetric Anderson model 
 at $U/{\Gamma}=3.5$ ( dashed line ) and  
$U/{\Gamma}=5.0$ ( solid line ).  
 }
\end{center}
\end{figure}
Figure 4.4 shows the spectral function with the self-energy  
 up to the fourth-order for equilibrium and zero temperature.
With strengthening $U,$ two-side narrow peaks come to occur 
 in the vicinity of $E=$ ${\pm}U/2$ in addition to the Kondo peak. 
At $U$ large enough, the Kondo peak becomes very acute 
and two-side narrow peaks rise higher and sharpen; 
the energy levels for the atomic limit are  produced distinctly.  
As before, the fourth-order self-energy has the same but narrow curves 
 as functions of energy with those of the second-order and 
those curves make the peaks at $E=$ ${\pm}U/2$. 

For the present approximation up to the fourth-order, 
the Kondo peak at $E=0$ reaches 
the unitarity limit and the charge, ${\langle}n{\rangle}$ 
corresponds to $1/2,$ that is, 
the Friedel sum rule is correctly satisfied[50]:  
\begin{eqnarray}
{\rho}(E_f)={\sin}^2({\pi}{\langle}n{\rangle})/
{\pi}{\Gamma}. 
\end{eqnarray}
where ${\rho}(E_f)$ is the local density of states at 
the Fermi energy. 

Here, the discussions should be made on the ranges of $U$ in which the 
present approximation up to the fourth-order stands. 
From the results,  it is found that the approximation within  
the  fourth-order holds up to $U/{\Gamma}$
${\sim}5.0$ and is beyond the validity for $U/{\Gamma}$$>{\sim}6.0$.  
 In such a case, the higher-order terms are required.  

 \ \ 

\subsection{For Self-Energy up to Fourth-Order for Nonequilibrium }

 \ \ 

\begin{figure}[ht]
\includegraphics[width=8.0cm]{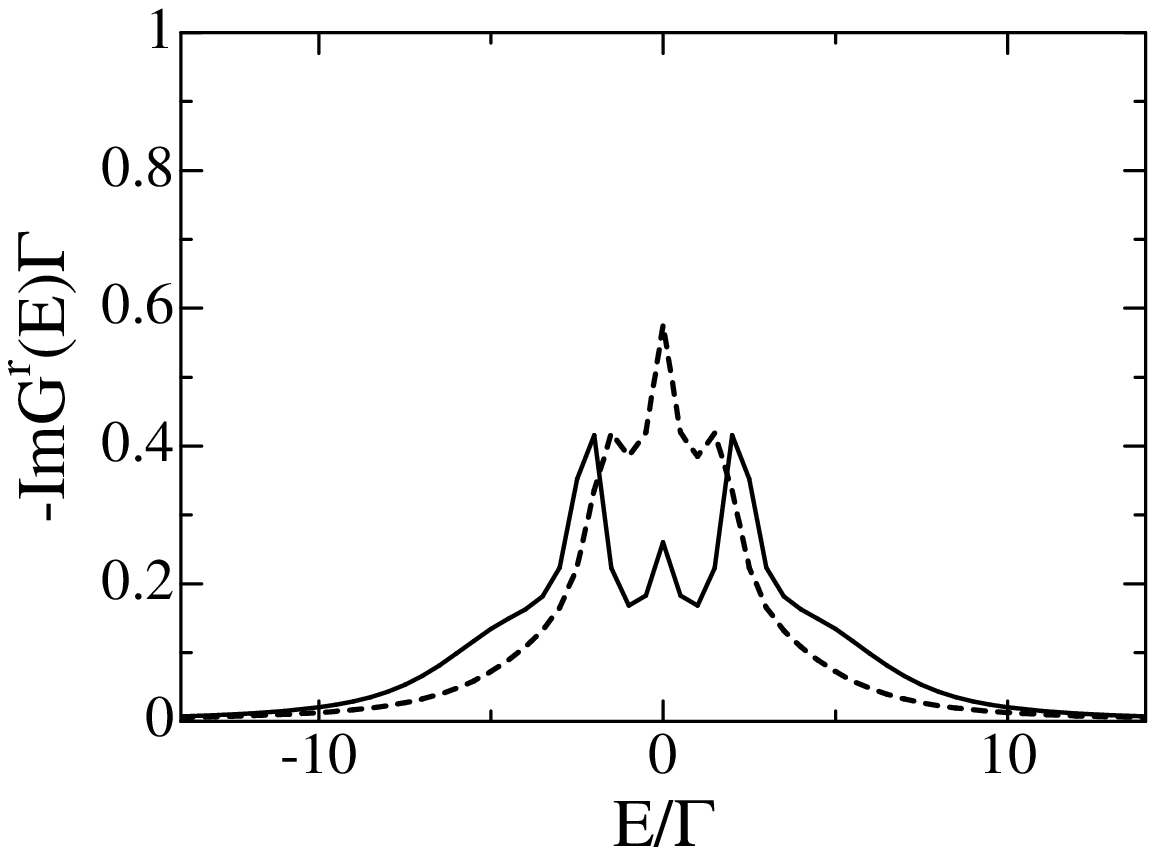}
\includegraphics[width=8.0cm]{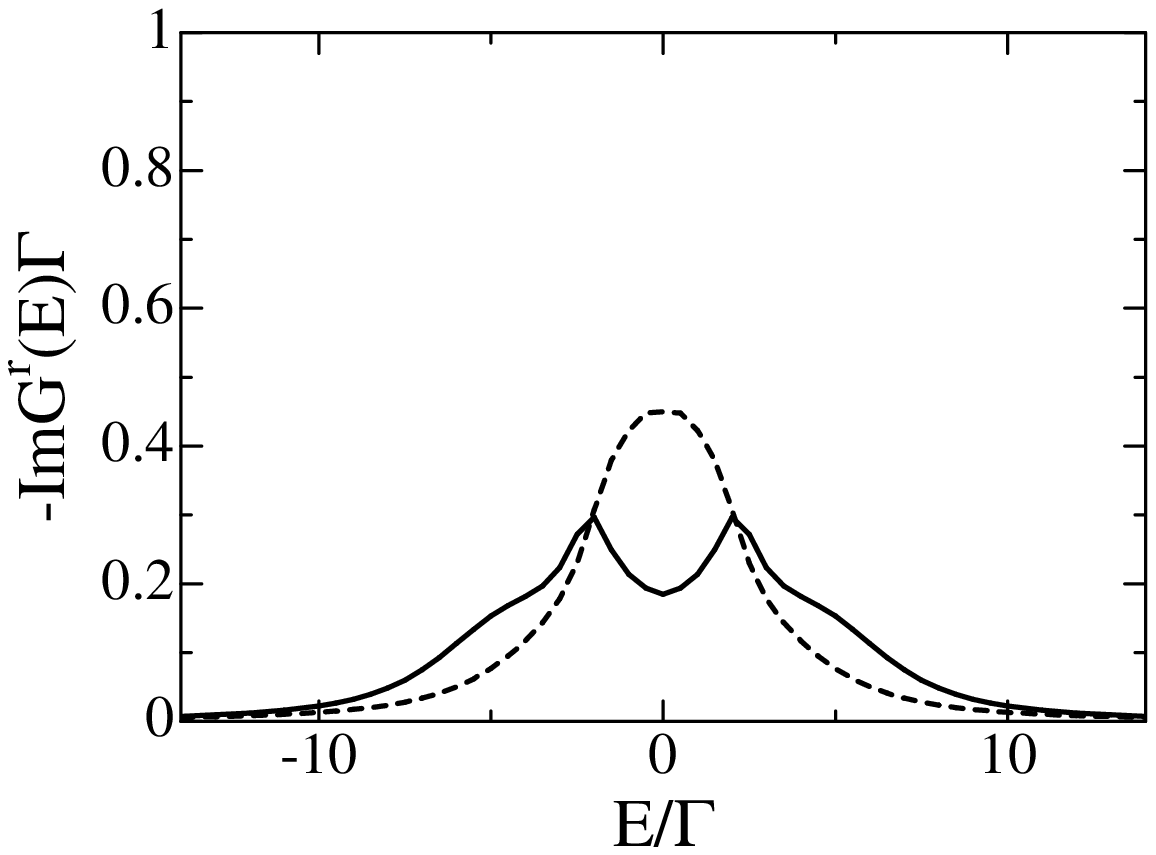}
\caption{
The spectral functions with  self-energy  up to the fourth-order  
at $eV/{\Gamma}=0.5$ ( Left ) and $eV/{\Gamma}=1.0$ ( Right )
for the symmetric Anderson model.  
$U/{\Gamma}=3.5$ ( dashed line ) and 
$U/{\Gamma}=5.0$ (  solid line ).
 }
\end{figure}
Next, the results for nonequilibrium and zero temperature are 
shown. 
The expression for the Friedel sum rule, Eq. ( 4.1 ) 
does not stand for nonequilibrium,  since the charge cannot be 
expressed with respect to  the local density of states.
 All the same, the Kondo peak reaches the unitarity limit 
and ${\langle}n{\rangle}=1/2$ in the symmetric and noninteracting case.

The spectral functions with the self-energy up to the fourth-order 
 are plotted for $eV/{\Gamma}=0.5$ and $eV/{\Gamma}=1.0$ 
in Fig. 4.5, respectively. 
The Kondo temperatures lessen with the rise of $U$,  
 as in general known. Approximately[49], 
\begin{eqnarray}
k_BT_K{\sim} \sqrt{ \frac{U{\Gamma}}{2} }
e^{-{\pi}U/8{\Gamma}+{\pi}{\Gamma}/2U}, 
\end{eqnarray}
As the estimation from Eq. ( 4.2 ), 
$k_BT_K/$${\Gamma}$ ${\sim}$$0.5$ for $U/$${\Gamma}$
$=3.5$ and $k_BT_K/$${\Gamma}$ ${\sim}$$0.3$ for $U/$${\Gamma}$$=5.0$.   
When $U$ is strengthened and $eV$ exceeds $k_BT_K$,  
the Kondo peak for $eV/{\Gamma}=0.5$ falls in and   
instead, the two-side narrow peaks remain to sharpen in the vicinity 
of $E=$ ${\pm}U/2$. For $eV/{\Gamma}=1.0,$ 
the Kondo peak becomes broad and disappears for $U$ large enough. 
The two-side peaks is generated small in the vicinity of $E=$ ${\pm}U/2$. 
The Kondo resonance is quite broken for bias voltage exceeding the Kondo temperatures;  
this accords with the experimental results of two terminal systems  
that the Kondo effect is suppressed when source-drain bias voltage is 
comparable to or exceeds the Kondo temperatures,  
 $eV  {\ge} k_BT_K$.[32,33]

For $eV/{\Gamma}>{\sim}2.0,$ the Kondo peak does not lower  
even when $eV$ is much larger than $k_BT_K$. 
The perturbative expansion is hard to converge 
on account of the imaginary part of the self-energy 
for $eV/{\Gamma}>{\sim}2.0,$ as described before; thereby, 
the higher-order contribution is probably 
needed for high voltage. 
Additionally, for high bias voltage, 
the picture of the nonequilibrium state represented 
as the superposition of the two leads is hard 
to stand. 

 \ \ 

We have to consider not only the Kondo effect but also 
the nonequilibrium state caused by bias voltage. 
In the present work, nonequilibrium state is represented 
as the superposition of the two leads. 
As is known, this works well as expression 
of nonequilibrium state.  
In this connection,  
the Khlus-Lesovik formula[11], the expression for current noise 
is drawn from the present picture[10] and is quantitatively   
consistent with experiments in ballistic systems. 
Thereby, the present picture can be valid as 
description of nonequilibrium state.   
The effective Fermi distribution function, Eq. ( 3.8 ) 
is found qualitatively similar to that for finite temperatures 
$f_{ T}(E)$. For example, for finite temperatures $T{\neq}0$, 
\begin{eqnarray}
f_{T}(E)[1-f_{ T}(E)]{\neq}0, \nonumber 
\end{eqnarray}
it is indicated that there are thermal charge fluctuations.
In case of the effective Fermi distribution function, 
\begin{eqnarray}
f_{\rm eff}(E)[1-f_{\rm eff}(E)]{\neq}0 \nonumber  
\end{eqnarray}
for finite voltage $eV{\neq}0$. 
From this analogy in the Fermi distribution function, 
it is inferred that 
there are nonequilibrium fluctuations similar to  
 thermal fluctuations. Because of the effective Fermi 
distribution function, not only for the second-order 
but also for the fourth-order, the Kondo resonance is destroyed, 
qualitatively the same as for finite temperatures,[51] 
as is observed in experiments of two terminal systems.
This is because the energy states in the leads 
are not discrete but continuous. 
The present results indicate that the Kondo resonance splitting 
due to bias voltage does not take place 
for simple two terminal systems. 

 \ \ 

Over a decade, it has been expected that 
the Kondo resonance splits by bias voltage 
and work as two channels for two channel Kondo effect, 
so that many endeavours have been made to seek 
the Kondo peak splitting. 
Recently, it has been reported that 
the two channel Kondo effect is observed 
in quantum dots system.[52] On the experiment, 
it is not observed for two infinite leads system 
which has continuous energy state. 
That occurs for the system that a large dot 
with closely discrete energy levels is added 
to two infinite leads system where a large dot works  
as finite size reservoir and two infinite leads system 
 acts as an infinite reservoir. 
The observed differential conductance 
depends on bias voltage and temperatures; 
this characteristic shows the functions 
of bias voltage and temperatures 
for two channel Kondo effect,   
derived by the theoretical work 
with conformal field theory.[20-22]   

 \ \

\chapter{Summary }

 \ \ 

In the present work, the solution is proposed 
 so as to make progress in the Schwinger-Keldysh 
formalism of nonequilibrium perturbation theory. 
Using the solution, the expressions for self-energies 
 are derived from the perturbative expansion 
in real-time through the Wick's theorem. 
On derivation, the expressions drawn via perturbative 
expansions can be taken for sum of 
retarded and advanced terms, and 
it is demonstrated that the advanced term in  
the retarded self-energy and the retarded term 
in the advanced self-energy vanish.  
As the consequence, the retarded self-energy is obtained 
as retarded function in time and 
the advanced self-energy is expressed as 
advanced function.  
These expressions for self-energies agree with 
those acquired by executing perturbative expansion in imaginary-time 
and analytical continuity;  
it is proven that the method of nonequilibrium perturbation theory 
can be linked with that of perturbation theory of thermal Green's functions.   
It is indicated that the Schwinger-Keldysh formalism works 
appropriately and that the formalism can be united with the 
other valid methods. The Schwinger-Keldysh formalism should be
 clarified and generalized furthermore. 

 \ \ 

As the numerical results, 
the Kondo peak disappears as bias voltage exceeding the Kondo temperatures.  
Because of the analogy of the effective Fermi distribution function 
for nonequilibrium with that for finite temperatures, 
the present result is qualitatively similar to that for finite temperatures.  
This characteristic appears in the experiments of two terminal systems. 
Consequently, the split of the Kondo peak by finite 
bias voltage may not occur in simple two terminal systems 
by reason of that the leads have continuous energy state. 

 \ \ 

The nonequilibrium systems in the presence of bias voltage  
are too complicated and of difficult access.  
Apart from the nonequilibrum perturbation theory 
as the present work, 
the statistical physics approach to the 
nonequilibrium systems is well-known 
as nonequilibrium statistical mechanics 
 and nonequilibrium statistical thermodynamics 
by Zubarev; nonequilibrium statistical 
operator has also been studied 
as the generalized Gibbs 
operator ( Gibbs operator is given by 
Eq. ( 2.12 ) ).[53]  
To the problem on the nonequilibrium systems 
by bias voltage, 
various approaches of both the statistical physics 
and phenomenology are also required.   
As for the Kondo physics, 
the various problems about  
the Kondo effect in nonequilibrium systems  
 and the multichannel Kondo effect   
remain to be clarified more. 

\chapter{Appendix }
\section{Appendix A \\ Fourth-Order Contribution to Self-Energy}
\makeatletter
 \renewcommand{\theequation}{%
  \thechapter.\arabic{equation}}
 \@addtoreset{equation}{section}
\makeatother
\renewcommand{\theequation}{
  \Alph{section}.\arabic{equation}}

The twelve terms for the fourth-order contribution  
 can be divided into four groups, each of which is composed  
of three terms. The four groups are brought from diagrams 
 denoted in  Fig. 3.3 (a)-(c), (d)-(f), (g)-(i), 
and (j)-(l), respectively. 
The terms for the diagrams illustrated 
in Fig3.3 (a) and (b) are equivalent 
except for the spin indices and expressed by 
\begin{eqnarray}
 {\Sigma}^{r(4)}_{a, b}(E)
&=&U^4\int^{\infty}_{0} {dt_{1}}\int^{\infty}_{-\infty}
{dt_{2}}\int^{\infty}_{-\infty}{dt_{3}}\, e^{iEt_1} \nonumber \\ 
  \ \  & & {\times} \left[ 
\begin{array}
{llll}
 g^<(t_1)
 g^<(t_1-t_2-t_3)
 g^>(-t_1+t_2+t_3) \nonumber \\ 
 -  g^>(t_1)
 g^>(t_1-t_2-t_3)
 g^<(-t_1+t_2+t_3) \nonumber \\ 
\end{array} 
\right] \nonumber \\
  \ \  & & {\times}  
\left[ \begin{array}
{llll}
 g^{\pm}(t_2)
 g^<(-t_2)  
 + g^<(t_2)
 g^{\pm}(-t_2) 
\end{array} 
\right] \nonumber \\
  \ \  & & {\times}\left[ \begin{array}
{llll}
 g^{\pm}(t_3)
 g^<(-t_3)  
 + g^<(t_3)
 g^{\pm}(-t_3)  
\end{array} 
\right],  
\end{eqnarray}
\begin{eqnarray}
 {\Sigma}^{a(4)}_{a, b}(E)
&=&U^4\int^{0}_{-\infty} {dt_{1}}\int^{\infty}_{-\infty}
{dt_{2}}\int^{\infty}_{-\infty}{dt_{3}}\, e^{iEt_1}
\nonumber \\   \ \  & & {\times}\left[ 
\begin{array}
{llll}
 g^>(t_1)
 g^>(t_1-t_2-t_3)
 g^<(-t_1+t_2+t_3) \nonumber \\ 
-g^<(t_1)
 g^<(t_1-t_2-t_3)
 g^>(-t_1+t_2+t_3) \nonumber \\ 
\end{array} 
\right] \\
 \ \     & &{\times} 
\left[ \begin{array}
{llll}
 g^{\pm}(t_2)
 g^<(-t_2)  
 + g^<(t_2)
 g^{\pm}(-t_2) 
\end{array} 
\right]   \nonumber \\ 
\ \    & &{\times}\left[ \begin{array}
{llll}
 g^{\pm}(t_3)
 g^<(-t_3)  
 + g^<(t_3)
 g^{\pm}(-t_3)  
\end{array} 
\right]. 
\end{eqnarray}
Additionally, Figure 3.3(c) shows the diagram 
for the following terms:  
\begin{eqnarray}
 {\Sigma}^{r(4)}_{c}(E)
&=&U^4\int^{\infty}_{0} {dt_{1}}\int^{\infty}_{-\infty}
{dt_{2}}\int^{\infty}_{-\infty}{dt_{3}}\, e^{iEt_1}
\nonumber \\ & & {\times} \left[ 
\begin{array}
{llll}
 g^>(-t_1)
 g^<(t_1-t_2-t_3)
 g^<(t_1-t_2-t_3) \nonumber \\ 
 -  g^<(-t_1)
 g^>(t_1-t_2-t_3)
 g^>(t_1-t_2-t_3) \nonumber \\ 
\end{array} 
\right] \\
  \ \  & & {\times} 
\left[ \begin{array}
{llll}
 g^{\pm}(t_2)
 g^>(t_2)  
 + g^<(t_2)
 g^{\pm}(t_2)  \end{array} 
\right]  \nonumber \\
  \ \  & & {\times}
\left[ \begin{array}
{llll}
 g^{\pm}(t_3)
 g^>(t_3)  
 + g^<(t_3)
 g^{\pm}(t_3)  \end{array} 
\right], 
\end{eqnarray}
\begin{eqnarray}
 {\Sigma}^{a(4)}_{c}(E)
&=&U^4\int^{0}_{-\infty} {dt_{1}}\int^{\infty}_{-\infty}
{dt_{2}}\int^{\infty}_{-\infty}{dt_{3}}\, e^{iEt_1}
\nonumber \\  & & {\times} \left[ 
\begin{array}
{llll}
 g^<(-t_1)
 g^>(t_1-t_2-t_3)
 g^>(t_1-t_2-t_3) \nonumber \\ 
 -g^>(-t_1)
 g^<(t_1-t_2-t_3)
 g^<(t_1-t_2-t_3) \nonumber \\ 
\end{array} 
\right] \\
 \ \   & & {\times} 
\left[ \begin{array}
{llll}
 g^{\pm}(t_2)
 g^>(t_2)  
 + g^<(t_2)
 g^{\pm}(t_2)  \end{array} 
\right] \nonumber \\
 \ \  & & {\times} 
\left[ \begin{array}
{llll}
 g^{\pm}(t_3)
 g^>(t_3)  
 + g^<(t_3)
 g^{\pm}(t_3)  \end{array} 
\right]. 
\end{eqnarray}
Next, the terms brought from diagram  
 in Fig. 3.3(d) are expressed by 
\begin{eqnarray}
 {\Sigma}^{r(4)}_{d}(E)
&=&U^4\int^{\infty}_{0} {dt_{1}}\int^{\infty}_{-\infty}
{dt_{2}}\int^{\infty}_{-\infty}{dt_{3}}\, e^{iEt_1}
\nonumber \\   & & \ \  {\times} \left[ 
\begin{array}
{llll}
 g^>(t_1-t_3)
 g^>(t_1-t_2) g^<(t_2-t_1)
 \nonumber \\ 
 -  g^<(t_1-t_3) g^<(t_1-t_2)
 g^>(t_2-t_1)
 \nonumber \\ 
\end{array} 
\right] \\
 & & \ \  {\times}
 \,g^{\pm}(t_2)\,{\rm sgn}(t_3)
\left[ \begin{array}
{llll}
 g^>(-t_2+t_3) g^>(t_3)
 g^<(-t_3)   \\ 
 -   g^<(-t_2+t_3)g^<(t_3)
 g^>(-t_3)
\end{array}
\right], 
\end{eqnarray}
\begin{eqnarray}
 {\Sigma}^{a(4)}_{d}(E)
&=&U^4\int^{0}_{-\infty} {dt_{1}}\int^{\infty}_{-\infty}
{dt_{2}}\int^{\infty}_{-\infty}{dt_{3}}\, e^{iEt_1}
\nonumber \\ & & {\times} \left[ 
\begin{array}
{llll}
g^<(t_1-t_3) g^<(t_1-t_2)
 g^>(t_2-t_1)
 \nonumber \\ 
-g^>(t_1-t_3)
 g^>(t_1-t_2) g^<(t_2-t_1)
 \nonumber \\ 
\end{array} 
\right] \\
& & {\times} \,g^{\pm}(t_2)\,{\rm sgn}(t_3)
\left[ 
\begin{array}
{llll}
 g^>(-t_2+t_3) g^>(t_3)
 g^<(-t_3)    \\ 
 -   g^<(-t_2+t_3)g^<(t_3)
 g^>(-t_3)
\end{array}
\right]. 
\end{eqnarray}
The terms for diagram in Fig. 3.3(e) are written by 
\begin{eqnarray}
 {\Sigma}^{r(4)}_{e}(E)
&=&U^4\int^{\infty}_{0} {dt_{1}}\int^{\infty}_{-\infty}
{dt_{2}}\int^{\infty}_{-\infty}{dt_{3}}\, e^{iEt_1}
\nonumber \\ & &   {\times}\left[ 
\begin{array}
{llll}
 g^>(t_1-t_2)
 g^>(t_1-t_2) g^<(t_3-t_1)
 \nonumber \\ 
 -  g^<(t_1-t_2) g^<(t_1-t_2)
 g^>(t_3-t_1)
 \nonumber \\ 
\end{array} 
\right] \\
 & &  {\times} \,g^{\pm}(t_2)\,{\rm sgn}(t_3)
\left[ 
\begin{array}
{llll}
 g^>(t_2-t_3) g^>(-t_3)
 g^<(t_3)    \\ 
 -   g^<(t_2-t_3)g^<(-t_3)
 g^>(t_3) \\
\end{array}
\right], 
\end{eqnarray}
\begin{eqnarray}
 {\Sigma}^{a(4)}_{e}(E)
&=&U^4\int^{0}_{-\infty} {dt_{1}}\int^{\infty}_{-\infty}
{dt_{2}}\int^{\infty}_{-\infty}{dt_{3}}\, e^{iEt_1} 
\nonumber \\ & & {\times} \left[ 
\begin{array}
{llll}
 g^<(t_1-t_2) g^<(t_1-t_2)
 g^>(t_3-t_1)
 \nonumber \\ 
- g^>(t_1-t_2)
 g^>(t_1-t_2) g^<(t_3-t_1)
 \nonumber \\ 
\end{array} 
\right] \\
  & & {\times} \,g^{\pm}(t_2)\,{\rm sgn}(t_3)
\left[ 
\begin{array}
{llll}
 g^>(t_2-t_3) g^>(-t_3)
 g^<(t_3)    \\ 
 -   g^<(t_2-t_3)g^<(-t_3)
 g^>(t_3)\\
\end{array}
\right].  
\end{eqnarray}
In addition, Figure 3.3(f) denotes the diagram for the 
following terms: 
\begin{eqnarray}
 {\Sigma}^{r(4)}_{f}(E)
&=&U^4\int^{\infty}_{0} {dt_{1}}\int^{\infty}_{-\infty}
{dt_{2}}\int^{\infty}_{-\infty}{dt_{3}}\, e^{iEt_1}
\nonumber \\ & & {\times} \left[ 
\begin{array}
{llll}
 g^>(t_1-t_3)
 g^>(t_1-t_2) g^<(t_2-t_1)
 \nonumber \\ 
 -  g^<(t_1-t_3) g^<(t_1-t_2)
 g^>(t_2-t_1)
 \nonumber \\ 
\end{array} 
\right] \\
    & &{\times} \,g^{\pm}(-t_2)\,{\rm sgn}(t_3)
\left[ 
\begin{array}
{llll}
 g^<(t_3)g^<(t_3)
 g^>(t_2-t_3)   \\ 
- g^>(t_3) g^>(t_3)
 g^<(t_2-t_3)  
\end{array}
\right], 
\end{eqnarray}
\begin{eqnarray}
 {\Sigma}^{a(4)}_{f}(E)
&=&U^4\int^{0}_{-\infty} {dt_{1}}\int^{\infty}_{-\infty}
{dt_{2}}\int^{\infty}_{-\infty}{dt_{3}}\, e^{iEt_1} 
\nonumber \\
& & {\times}
\left[ 
\begin{array}
{llll}
 g^<(t_1-t_3) g^<(t_1-t_2)
 g^>(t_2-t_1)
 \nonumber \\ 
- g^>(t_1-t_3)
 g^>(t_1-t_2) g^<(t_2-t_1)
 \nonumber \\ 
\end{array} 
\right] \\
   & &{\times} \,g^{\pm}(-t_2)\,{\rm sgn}(t_3)
\left[ 
\begin{array}
{llll}
 g^<(t_3)g^<(t_3)
 g^>(t_2-t_3)   \\ 
  - g^>(t_3) g^>(t_3)
 g^<(t_2-t_3)  
\end{array}
\right]. 
\end{eqnarray}
Next, the terms formulated from diagram   
illustrated in Fig. 3.3(g) are expressed by 
\begin{eqnarray}
 {\Sigma}^{r(4)}_{g}(E)
&=&U^4\int^{\infty}_{0} {dt_{1}}\int^{\infty}_{-\infty}
{dt_{2}}\int^{\infty}_{-\infty}{dt_{3}}\, e^{iEt_1}
\nonumber \\
& & {\times}
\left[ 
\begin{array}
{llll}
 g^>(t_1)
 g^>(t_1-t_2-t_3)
 g^<(t_2-t_1) \nonumber \\ 
 -  g^<(t_1)
 g^<(t_1-t_2-t_3)
 g^>(t_2-t_1) \nonumber \\ 
\end{array} 
\right] \\
 \ \   & &{\times} \,g^{\pm}(-t_2)\,{\rm sgn}(t_3)
\left[ 
\begin{array}
{llll}
  g^>(t_2+t_3)g^>(t_3)
 g^<(-t_3)   \\ 
 -   g^<(t_2+t_3)g^<(t_3)
 g^>(-t_3)
\end{array}
\right], \nonumber \\
\end{eqnarray}
\begin{eqnarray}
 {\Sigma}^{a(4)}_{g}(E)
&=&U^4\int^{0}_{-\infty} {dt_{1}}\int^{\infty}_{-\infty}
{dt_{2}}\int^{\infty}_{-\infty}{dt_{3}}\, e^{iEt_1}
\nonumber \\
& & {\times}
\left[ 
\begin{array}
{llll}
 g^<(t_1)
 g^<(t_1-t_2-t_3)
 g^>(t_2-t_1) \nonumber \\ 
- g^>(t_1)
 g^>(t_1-t_2-t_3)
 g^<(t_2-t_1) \nonumber \\ 
\end{array} 
\right] \\
 \ \   & & {\times} \,g^{\pm}(-t_2)\,{\rm sgn}(t_3)
\left[ 
\begin{array}
{llll}
 g^>(t_2+t_3) g^>(t_3)
 g^<(-t_3)   \\ 
 -  g^<(t_2+t_3)g^<(t_3)
 g^>(-t_3)
\end{array}
\right]. \nonumber \\
\end{eqnarray}
Figure 3.3(h) illustrates the diagram for 
the following terms:
\begin{eqnarray}
 {\Sigma}^{r(4)}_{h}(E)
&=&U^4\int^{\infty}_{0} {dt_{1}}\int^{\infty}_{-\infty}
{dt_{2}}\int^{\infty}_{-\infty}{dt_{3}}\, e^{iEt_1}
\nonumber \\
& & {\times}
\left[ 
\begin{array}
{llll}
 g^<(t_1)
 g^<(t_1-t_2-t_3)
 g^>(t_2-t_1) \nonumber \\ 
 -  g^>(t_1)
 g^>(t_1-t_2-t_3)
 g^<(t_2-t_1) \nonumber \\ 
\end{array} 
\right] \\
 \ \   & & {\times} \,g^{\pm}(t_2)\,{\rm sgn}(t_3)
\left[ 
\begin{array}
{llll}
 g^>(t_3)
 g^>(t_3)
 g^<(-t_2-t_3)    \\ 
 -  g^<(t_3)
 g^<(t_3)
 g^>(-t_2-t_3)  
\end{array}
\right], \nonumber \\
\end{eqnarray}
\begin{eqnarray}
 {\Sigma}^{a(4)}_{h}(E)
&=&U^4\int^{0}_{-\infty} {dt_{1}}\int^{\infty}_{-\infty}
{dt_{2}}\int^{\infty}_{-\infty}{dt_{3}}\, e^{iEt_1}
\nonumber \\
& & {\times}
\left[ 
\begin{array}
{llll}
 g^>(t_1)
 g^>(t_1-t_2-t_3)
 g^<(t_2-t_1) \nonumber \\ 
-g^<(t_1)
 g^<(t_1-t_2-t_3)
 g^>(t_2-t_1) \nonumber \\ 
\end{array} 
\right] \\
 \ \   & & {\times} \,g^{\pm}(t_2)\,{\rm sgn}(t_3)
\left[ 
\begin{array}
{llll}
 g^>(t_3)
 g^>(t_3)
 g^<(-t_2-t_3)    \\ 
 -  g^<(t_3)
 g^<(t_3)
 g^>(-t_2-t_3)  
\end{array}
\right]. \nonumber \\
\end{eqnarray}
Besides, the terms formulated from the diagram in Fig. 3.3(i) 
 are written by 
\begin{eqnarray}
 {\Sigma}^{r(4)}_{i}(E)
&=&U^4\int^{\infty}_{0} {dt_{1}}\int^{\infty}_{-\infty}
{dt_{2}}\int^{\infty}_{-\infty}{dt_{3}}\, e^{iEt_1}
\nonumber \\
& & {\times}
\left[ 
\begin{array}
{llll}
 g^>(-t_1)
 g^<(t_1-t_2-t_3)
 g^<(t_1-t_2) \nonumber \\ 
 -  g^<(-t_1)
 g^>(t_1-t_2-t_3)
 g^>(t_1-t_2) \nonumber \\ 
\end{array} 
\right] \\
 \ \   & & {\times} \,g^{\pm}(t_2)\,{\rm sgn}(t_3)
\left[ 
\begin{array}
{llll}
  g^>(t_2+t_3)g^>(t_3)
 g^<(-t_3)   \\ 
 - g^<(t_2+t_3) g^<(t_3)
 g^>(-t_3)
\end{array}
\right], \nonumber \\
\end{eqnarray}
\begin{eqnarray}
 {\Sigma}^{a(4)}_{i}(E)
&=&U^4\int^{0}_{-\infty} {dt_{1}}\int^{\infty}_{-\infty}
{dt_{2}}\int^{\infty}_{-\infty}{dt_{3}}\, e^{iEt_1}
\nonumber \\
& & {\times}
\left[ 
\begin{array}
{llll}
 g^<(-t_1)
 g^>(t_1-t_2-t_3)
 g^>(t_1-t_2) \nonumber \\ 
- g^>(-t_1)
 g^<(t_1-t_2-t_3)
 g^<(t_1-t_2) \nonumber \\ 
\end{array} 
\right] \\
 \ \   & & {\times} \,g^{\pm}(t_2)\,{\rm sgn}(t_3)
\left[ 
\begin{array}
{llll}
  g^>(t_2+t_3)g^>(t_3)
 g^<(-t_3)    \\ 
 -  g^<(t_2+t_3)  g^<(t_3)
 g^>(-t_3)
\end{array}
\right].\nonumber \\
\end{eqnarray}
Next, the terms for diagrams denoted in Figs. 3.3 (j) and (k)  
are equivalent except for the spin indices and written by 
\begin{eqnarray}
 {\Sigma}^{r(4)}_{j, k}(E)
&=&U^4\int^{\infty}_{0} {dt_{1}}\int^{\infty}_{-\infty}
{dt_{2}}\int^{\infty}_{-\infty}{dt_{3}}\, e^{iEt_1}
\nonumber \\
& & {\times}
\left[ 
\begin{array}
{llll}
 g^>(t_1)
 g^<(-t_1)
 g^>(t_1-t_2-t_3) \nonumber \\ 
 -  g^<(t_1)
 g^>(-t_1)
 g^<(t_1-t_2-t_3) \nonumber \\ 
\end{array} 
\right] \\
 \ \   & & {\times} \,g^{\pm}(t_2)
\left[ 
\begin{array}
{llll}
\ \  g^{\pm}(t_3)
 g^>(t_3)
 g^<(-t_3)     \\
 +  g^<(t_3)
 g^{\pm}(t_3)
 g^>(-t_3)   \\
 +  g^<(t_3)
 g^>(t_3)
 g^{\pm}(-t_3)  \\ 
\end{array}
\right], 
\end{eqnarray}
\begin{eqnarray}
 {\Sigma}^{a(4)}_{j, k}(E)
&=&U^4\int^{0}_{-\infty} {dt_{1}}\int^{\infty}_{-\infty}
{dt_{2}}\int^{\infty}_{-\infty}{dt_{3}}\, e^{iEt_1}
\nonumber \\
& & {\times}
\left[ 
\begin{array}
{llll}
  g^<(t_1)
 g^>(-t_1)
 g^<(t_1-t_2-t_3) \nonumber \\ 
- g^>(t_1)
 g^<(-t_1)
 g^>(t_1-t_2-t_3) \nonumber \\ 
\end{array} 
\right] \\
 \ \   & & {\times} \,g^{\pm}(t_2)
\left[ 
\begin{array}
{llll}
\ \  g^{\pm}(t_3)
 g^>(t_3)
 g^<(-t_3)   \\
 +  g^<(t_3)
 g^{\pm}(t_3)
 g^>(-t_3)  \\ 
 +  g^<(t_3)
 g^>(t_3)
 g^{\pm}(-t_3)  \\ 
\end{array}
\right]. 
\end{eqnarray}
In addition, the terms for diagram illustrated in Fig. 3.3(l)  
are expressed by 
\begin{eqnarray}
 {\Sigma}^{r(4)}_{l}(E)
&=&U^4\int^{\infty}_{0} {dt_{1}}\int^{\infty}_{-\infty}
{dt_{2}}\int^{\infty}_{-\infty}{dt_{3}}\, e^{iEt_1}
\nonumber \\
& & {\times}
\left[ 
\begin{array}
{llll}
 g^>(t_1)
 g^>(t_1)
 g^<(-t_1+t_2+t_3) \nonumber \\ 
 -  g^<(t_1)
 g^<(t_1)
 g^>(-t_1+t_2+t_3) \nonumber \\ 
\end{array} 
\right] \nonumber \\
 \ \   & & {\times} \,g^{\pm}(-t_2)
\left[ 
\begin{array}
{llll}
\ \ g^{\pm}(-t_3)
 g^>(-t_3) g^<(t_3)  \\
 +  g^<(-t_3) g^{\pm}(-t_3)
 g^>(t_3) \\  +  g^<(-t_3)
 g^>(-t_3)
 g^{\pm}(t_3)   \end{array} 
\right],    
\end{eqnarray}
\begin{eqnarray}
 {\Sigma}^{a(4)}_{l}(E)
&=&U^4\int^{0}_{-\infty} {dt_{1}}\int^{\infty}_{-\infty}
{dt_{2}}\int^{\infty}_{-\infty}{dt_{3}}\, e^{iEt_1}
\nonumber \\
& & {\times}
\left[ 
\begin{array}
{llll}
 g^<(t_1)
 g^<(t_1)
 g^>(-t_1+t_2+t_3) \nonumber \\ 
-g^>(t_1)
 g^>(t_1)
 g^<(-t_1+t_2+t_3) \nonumber \\ 
\end{array} 
\right] \nonumber \\
 \ \   & & {\times} \,g^{\pm}(-t_2)
\left[ 
\begin{array}
{llll}
\ \ g^{\pm}(-t_3)
 g^>(-t_3) g^<(t_3) \\ 
 +  g^<(-t_3) g^{\pm}(-t_3)
 g^>(t_3) \\  +  g^<(-t_3)
 g^>(-t_3)
 g^{\pm}(t_3)   \end{array} 
\right]. 
\end{eqnarray}

 \ \ 

 \ \ 

\section{Appendix B \ \ \\
Expressions for Magnetization and Susceptibility }

 \ \

In the connection with the nonequilibrium state, 
the expressions for magnetization and 
susceptibility are written. Here, we consider the system 
where the magnetic field is applied 
to the impurity. 
There the Zeeman term of the impurity, $-BS_Z$
( $B$ is magnetic field  ) is added to the Anderson 
Hamiltonian. 
Magnetization for spin $1/2$ is written by 
\begin{eqnarray}
M={\langle}S_Z{\rangle}={\frac{1}{2}}
({\langle}\hat{n}_{d\uparrow}{\rangle}
-{\langle}\hat{n}_{d\downarrow}{\rangle})
={\frac{1}{4{\pi}i}}\int{dE}
[G^<_{\uparrow}(E)-G^<_{\downarrow}(E)], 
\end{eqnarray}
where 
\begin{eqnarray}
{\langle}\hat{n}_{d\uparrow}{\rangle}
={\frac{1}{2{\pi}i}}\int{dE}
G^<_{\uparrow}(E). \nonumber
\end{eqnarray}

For simplification, it is assumed that 
the system is noninteracting ( $U=0$ ) 
and has symmetries: ${\Gamma}_L={\Gamma}_R$, 
 $({\Gamma}_L+{\Gamma}_R)/2={\Gamma}$,   
 and the applied voltage: ${\mu}_L=eV/2$, ${\mu}_R=-V/2$.  

 \ \ 

The Fermi distribution function can be 
written by
\begin{eqnarray}
f(x)=(e^x+1)^{-1}={\frac{1}{2}}+{\frac{i}{2}}\tan\left({\frac{x}{2}}i\right).
\end{eqnarray}
From the formula of digamma function ${\psi}$, 
\begin{eqnarray}
-{\frac{1}{2{\pi}i}}
\left\{{\psi}\left[{\frac{1}{2}}+i{\frac{x}{2{\pi}}}\right]
-{\psi}\left[{\frac{1}{2}}-i{\frac{x}{2{\pi}}}\right]\right\}
={\frac{i}{2}}\tan\left({\frac{x}{2}}i\right).
\end{eqnarray}
Accordingly, the Fermi distribution function can be written 
in terms of digamma function ${\psi}$ by 
\begin{eqnarray}
f(x)={\frac{1}{2}}-{\frac{1}{2{\pi}i}}
\left\{{\psi}\left[{\frac{1}{2}}+i{\frac{x}{2{\pi}}}\right]
-{\psi}\left[{\frac{1}{2}}-i{\frac{x}{2{\pi}}}\right]\right\}. 
\end{eqnarray}
The charge is expressed in terms of the Fermi distribution function 
using the residue theorem for finite magnetic field by  
\begin{eqnarray}
{\langle}\hat{n}_{d{\uparrow}{\downarrow}}{\rangle}
=f\left( {\frac{{\pm}B-i{\Gamma}}{T}} \right), 
\end{eqnarray}
where $T$  is temperature. 
If the right-hand side of Eq. ( B.5 ) 
is replaced with Eq. ( B.4 ),  then, 
\begin{eqnarray}
{\langle}\hat{n}_{d{\uparrow}{\downarrow}}{\rangle}
&=&{\frac{1}{2}}-{\frac{1}{2{\pi}i}}
\left\{{\psi}\left[{\frac{1}{2}}+i{\frac{{\pm}B-i{\Gamma}}{2{\pi}T}}\right]
-{\psi}\left[{\frac{1}{2}}-i{\frac{{\pm}B-i{\Gamma}}{2{\pi}T}}\right]\right\}
 \nonumber \\
&=&{\frac{1}{2}}-{\frac{1}{\pi}}{\rm Im}{\psi}
\left[{\frac{1}{2}}+{\frac{{\pm}B+i{\Gamma}}{2{\pi}iT}}\right].  
\end{eqnarray}
Magnetization at equilibrium state ( $eV=0$ ), therefore,  reduces to 
\begin{eqnarray}
M(B)={\frac{1}{2}}
\left\{-{\frac{1}{\pi}}{\rm Im}{\psi}\left[{\frac{1}{2}}+
{\frac{B+i{\Gamma}}{2{\pi}iT}}\right]
+{\frac{1}{\pi}}{\rm Im}{\psi}\left[{\frac{1}{2}}+
{\frac{-B+i{\Gamma}}{2{\pi}iT}}\right]
\right\}.
\end{eqnarray}
For nonequilibrium state ( $eV{\neq}0$ ), 
the effective Fermi distribution function is obtained from 
Eq. ( 3.8 ) by 
\begin{eqnarray}
f_{\rm eff}(E)={\frac{f_L(E)+f_R(E)}{2}},  
\end{eqnarray}
 using this, then, magnetization at $eV{\neq}0$ is written by  
\begin{eqnarray}
M(B, eV) ={\frac{1}{4}}
\biggl\{  &-&{\frac{1}{\pi}}{\rm Im}{\psi}\left[{\frac{1}{2}}+
{\frac{B+eV/2+i{\Gamma}}{2{\pi}iT}}\right]
-{\frac{1}{\pi}}{\rm Im}{\psi}\left[{\frac{1}{2}}+
{\frac{B-eV/2+i{\Gamma}}{2{\pi}iT}}\right]   \nonumber \\
 &+&{\frac{1}{\pi}}{\rm Im}{\psi}\left[{\frac{1}{2}}+
{\frac{-B+eV/2+i{\Gamma}}{2{\pi}iT}}\right]
+{\frac{1}{\pi}}{\rm Im}{\psi}\left[{\frac{1}{2}}+
{\frac{-B-eV/2+i{\Gamma}}{2{\pi}iT}}\right]
\Biggr\}. \nonumber  \\
\end{eqnarray}
The expression for nonequilibrium state is written as the sum 
of term of each chemical potential in the leads. 

 \ \ 

When the limit is taken, the expressions 
are simplified as follows: 
in zero temperature limit, 
the expressions for magnetization $M$ and 
susceptibility ${\chi}$ at equilibrium state reduce to  
\begin{eqnarray}
M(B)&=&{\frac{1}{2\pi}}\left[{\arctan}\left({\frac{B}{{\Gamma}}}\right)
-{\arctan}\left({\frac{-B}{{\Gamma}}}\right)\right] \nonumber \\
&=&{\frac{1}{\pi}}{\arctan}\left({\frac{B}{{\Gamma}}}\right), \\
{\chi}(B)&=&{\frac{dM(B)}{dB}}={\frac{1}{\pi}}{\frac{{\Gamma}}{B^2+{\Gamma}^2.}} 
\end{eqnarray}
For nonequilibrium state( $eV{\neq}0$ ), 
\begin{eqnarray}
M(B, eV )&=&{\frac{1}{2{\pi}}}\left[
{\arctan}\left({\frac{B+eV/2}{{\Gamma}}}\right)
+{\arctan}\left({\frac{B-eV/2}{{\Gamma}}}\right)
\right], 
\nonumber \\
\end{eqnarray}
\begin{eqnarray}
{\chi}(B, eV)&=&{\frac{ {\Gamma}\left[
B^2+(eV/2)^2+{\Gamma}^2\right] }
{ {\pi} \left[(B+eV/2)^2+{\Gamma}^2\right]
\left[(B-eV/2)^2+{\Gamma}^2\right].}} 
\end{eqnarray}

 \ \ 


 \ \ 

In isolated limit ${\Gamma}{\rightarrow}0,$  
 that the connection of the quantum dot with leads vanishes,  
the expressions for magnetization for thermal equilibrium 
reduce to 
\begin{eqnarray}
M(B,T)&=&{\frac{1}{2}}\left[ {\frac {1}{e^{-B/T}+1}}
-{\frac {1}{e^{B/T}+1}} \right] \nonumber \\
&=&{\frac{1}{2}}{\rm tanh}\left({\frac{B}{2T}}\right), 
\end{eqnarray}
the Brillouin function as is known. In addition, 
susceptibility is also obtained by 
\begin{eqnarray}
{\chi}(B,T)=
{\frac{1}{4T}}{\rm sech}^2\left({\frac{B}{2T}}\right).
\end{eqnarray}

At nonequilibrium state,  
\begin{eqnarray}
M(B, T, eV)=
{\frac{1}{4}}\left[{\rm tanh}\left({\frac{B+eV/2}{2T}}\right)+
 {\rm tanh}\left({\frac{B-eV/2}{2T}}\right)\right],  
\end{eqnarray}
\begin{eqnarray}
{\chi}(B, T, eV)=
{\frac{1}{8T}}\left[{\rm sech}^2\left({\frac{B+eV/2}{2T}}\right)
+{\rm sech}^2\left({\frac{B-eV/2}{2T}}\right)\right].
\end{eqnarray}

 \ \ 

In consequence, the expressions at nonequilibrium state 
are gained as the sum 
of term of each chemical potential.

\chapter*{Acknowledgements}
\addcontentsline{toc}{chapter}{Acknowledgements}
The numerical calculations were executed 
at the Yukawa Institute Computer Facility. 
The multiple integrals were performed   
with the  computer subroutine, {\it MQFSRD} of NUMPAC.

\end{document}